\documentclass[10pt, doublecolumn]{IEEEtran}
\usepackage{graphicx}
\usepackage{caption}
\ifCLASSOPTIONcompsoc
  \usepackage[caption=false,font=normalsize,labelfont=sf,textfont=sf]{subfig}
\else
  \usepackage[caption=false,font=footnotesize]{subfig}
\fi
\usepackage{indentfirst}

\usepackage{amsmath}
\allowdisplaybreaks[4]
\usepackage{amssymb}
\usepackage{times}
\usepackage{mathtools}
\usepackage{psfrag}
\usepackage{cite}
\usepackage{lastpage}
\usepackage{fancyhdr}
\usepackage{color}
 \usepackage{amsthm}
\usepackage{bigints}
\usepackage{multirow}
\usepackage{booktabs}
\usepackage{threeparttable}

\newtheorem{Theorem}{Theorem}
\newtheorem{Lemma}{Lemma}
\newtheorem{Remark}{Remark}
\newtheorem{Corollary}{Corollary}

\newtheorem{lemma}[Lemma]{$\mathbf{Lemma}$}

\begin{document}
\title{Hybrid SIC Aided Hybrid NOMA: A New Approach For Improving Energy Efficiency}
\author{Yanshi Sun, \IEEEmembership{Member, IEEE}, Wei Cao, Ning Wang, Momiao Zhou, \IEEEmembership{Member, IEEE}, Zhiguo Ding, \IEEEmembership{Fellow, IEEE}
\thanks{
The work of Yanshi Sun and Wei Cao was supported in part by Hefei University of Technology's 
construction funds for the introduction of talents with funding number 13020-03712022011 
and the National Natural Science Foundation of China under Grant 62301208. 

Y. Sun, Wei Cao and M. Zhou  are  with the School of Computer Science and Information
Engineering, Hefei University of Technology, Hefei, 230009, China. (email: sys@hfut.edu.cn, caowei0115@163.com and mmzhou@hfut.edu.cn).

N. Wang is with School of Information and Control Engineering, China University of Mining and Technology, Xuzhou, 221116, China. (wangnsky@cumt.edu.cn).

Z. Ding is with Department of Computer and Information Engineering, Khalifa University, Abu Dhabi, UAE. (email: zhiguo.ding@ieee.org).
}\vspace{-2em}}
\maketitle
\begin{abstract}
Hybrid non-orthogonal multiple access (NOMA), which organically combines pure NOMA and conventional OMA, has recently received
significant attention to be a promising multiple access framework for future wireless communication networks.
However, most of the literatures on hybrid NOMA only consider fixed order of successive interference cancellation (SIC), namely FSIC, for the NOMA transmission phase of hybrid NOMA, resulting in limited performance.  Differently, this paper aims to reveal the potential of applying hybrid SIC (HSIC) to improve the energy efficiency
of hybrid NOMA. Specifically, a HSIC aided hybrid NOMA scheme is proposed, which can be treated as a simple add-on to the legacy orthogonal multiple access (OMA) based network. The proposed scheme offers some users (termed ``opportunistic users'') to have more chances to transmit by  transparently sharing legacy users' time slots. For a fair comparison, a power reducing coefficient $\beta$ is introduced to ensure that the energy consumption of the proposed scheme is less than conventional OMA. Given $\beta$, the probability for the event that the achievable rate of the proposed HSIC aided hybrid NOMA scheme cannot outperform its OMA counterpart is obtained in closed-form, by considering impact of user pairing.
{\color{black}Furthermore, asymptotic analysis shows that the aforementioned probability can approach zero in the high SNR regime under some given conditions, which are compositely  determined by the users' transmit powers, primary user's target data rate and $\beta$, indicating that the energy efficiency of the proposed scheme is almost surely higher than that of OMA for these given conditions.}
Numerical results are presented to verify the analysis and also demonstrate the benefit of applying HSIC compared to FSIC.
\end{abstract}
	
\section{Introduction}
Multiple access techniques plays a vital role in each generation wireless communication networks. However, most of existing networks adopt orthogonal multiple access (OMA) techniques. OMA techniques can well mitigate inter-user interferences, but has limited performance in terms of throughput and connectivity. Under this background, non-orthogonal multiple access (NOMA) techniques has recently received significant research interest \cite{you2021towards, 3GPPNOMAR16,liu2022evolution}. For example,
International Mobile Telecommunications (IMT)-2030 Framework issued by the International Telecommunication Union (ITU) has recognized NOMA as a crucial component for 6G \cite{recommendation2023framework}. The advantage of NOMA compared to OMA is that channel resources are encouraged to be shared among users, which may result in higher rate and larger connectivity. Besides, it has been shown by the literature that NOMA is compatible with many advancing techniques such as integrated sensing and communications (ISAC) \cite{sun2024study}, fluid antenna systems (FAS)\cite{new2023fluid}, reconfigurable intelligent surface (RIS) networks \cite{li2023achievable,zhu2020power}, semantic communications\cite{mu2023exploiting}, non-terrestrial networks (NTN) \cite{chu2020robust} and so on.

Due to the reason that existing mobile networks are developed based on OMA,
how to embed NOMA onto the legacy framework to deeply leverage the advantages of NOMA without
sophisticated disruptions to the existing network, becomes an important and practical issue.
To this end, the concept of hybrid NOMA has been proposed, which organically combines pure NOMA and existing OMA and
can be implemented as a simple add-on to the legacy OMA network \cite{Ding2019MEC, ding2024backcom}.
The key idea of hybrid NOMA is that some of the users in the network can not only  transmit signal by using its own
allocated channel resource block as in OMA, but also have other chances to transmit by using other users' channel
resource blocks by applying NOMA. Hybrid NOMA was initially proposed for the
mobile edge computing (MEC) scenario\cite{Ding2019MEC}. It has been shown that by applying hybrid NOMA
it is possible to offload more data with less energy consumption compared to OMA based schemes \cite{Ding2019MEC}.
{\color{black}Note that \cite{Ding2019MEC} only considered the application of hybrid NOMA to a two-user scenario, \cite{ding2022hybrid}
generalized the application of hybrid NOMA to the mutli-user scenario, showing that the energy consumption of
hybrid NOMA can outperform pure NOMA and OMA based schemes.
Particularly, in \cite{ding2022hybrid}, a multi-objective optimization problem was developed which aims to minimize the
users' energy consumption for MEC offloading, where a low-complexity Pareto-optimal solution for power allocation and time slot duration
configuration was obtained. \cite{liu2021latency} studied the latency minimization for MEC offloading by incorporating
three types of multiple access mehtods, namely hybrid NOMA, pure NOMA, and pure OMA.
The superior performance of hybrid NOMA in terms of energy efficiency has also been investigated in \cite{wei2022energy}.
Note that the aforementioned works mainly considered the scenario where hybrid NOMA is implemented over
a time division multiple access (TDMA) based legacy network. \cite{ding2024utilizing} further showed that
hybrid NOMA can also be applied to networks where the orthogonal resources are beams, to reduce energy consumption.
Particularly, a hybrid NOMA transmission strategy was desgined in \cite{ding2024utilizing} for near-feild scenarios,
where some users in the network can have additional transmisison opportunities by sharing
 beams which are preconfigured for legacy near-field users.
Note that existing works on hybrid NOMA mainly focus on uplink scenarios
\cite{Ding2019MEC,ding2024backcom,ding2022hybrid,liu2021latency,wei2022energy,ding2024utilizing}.
In \cite{ding2024design}, the application of hybrid NOMA in downlink single input single output (SISO) and multiple input
single output (MISO) scenarios were studied. It was shown by \cite{ding2024design} that hybrid NOMA always outperforms
OMA in terms of energy consumption, when the users' channel gains are ordered and the duration of time slots are the same.}
Tools from reinforcement learning has also been applied to
optimize resource allocation of hybrid NOMA for performance
enhancement in \cite{yu2022irs,wang2022reinforcement,chaieb2022deep}.

Note that for the NOMA transmission of hybrid NOMA,
successive interference cancellation (SIC) need to be carried out to mitigate inter-user interferences.
However, most of the literatures on hybrid NOMA only considered SIC methods where the decoding orders of users are fixed,
which are determined by ordering users according to their channel
gains \cite{higuchi2013non, gao2017theoretical,Xia2018outage} or quality of service (QoS)
requirements \cite{zhou2018state,Dhakal2019noma,Ding2019simple}.
{\color{black}Such SIC methods are termed fixed SIC (FSIC).
It has been shown by \cite{ding2021new, sun2021new, lu2022advanced,sun2023hybrid} hybrid SIC (HSIC) methods
which dynamically selecting the SIC orders of users by comprehensively judging the relationship of several factors
including channel gains, transmit power and QoS, can significantly improve the transmission robustness
for pure NOMA transmission in a single time slot. To elaborate on the difference between HSIC and FSIC methods,
consider a two-user scenario as in [11], including a primary user and secondary user,
where the secondary user can transmit signals not only in its own time slot but also in the primary user's time slot
by applying NOMA. In [11], for the NOMA transmission,
the secondary user's signal can only be decoded at the first stage of SIC, i.e., FSIC is applied.
In contrast, if HSIC is adopted, the secondary user's signal can be decoded either in the first or second stage of SIC,
depending on the channel gains, transmit power and QoS. As aforementioned,
although the benefit of HSIC in pure NOMA which focuses one single time slot has
been revealed by \cite{ding2021new, sun2021new, lu2022advanced,sun2023hybrid},
what role can HSIC play in hybrid NOMA which incorporates multiple time slots is still open.
Thus, it is natural to ask the following question: what benefit can HSIC bring to hybrid NOMA system,
which motivates the work of this paper.} 

This paper aims to investigate the benefit of applying HSIC to hybrid NOMA transmission to improve energy efficiency
from a probabilistic perspective. The main contributions of this paper are summarized as follows:
\begin{itemize}
  \item A novel hybrid SIC aided hybrid NOMA scheme is proposed based on a TDMA legacy network. The users are classified into two types, i.e., the legacy users and opportunistic users. The legacy users transmit signals by using their own slots. In contrast, besides their own slots, the opportunistic users have additional chances to transmit by sharing their paired legacy users' slots. For a fair comparison, a power reducing coefficient $\beta$ is introduced to ensure that the energy consumption of the proposed scheme is less than conventional OMA.
  \item Rigorous derivations are developed to obtain the closed-form expression for the probability (denoted by $\tilde{P}_n$) of the event that the achievable rate of the proposed HSIC aided hybrid NOMA scheme cannot outperform its OMA counterpart, for a given $\beta$ less than $\frac{1}{2}$. Based on the obtained results, asymptotic behavior of $\tilde{P}_n$ is analyzed, which shows that under some conditions, $\tilde{P}_n$ can approach zero for a given $\beta<\frac{1}{2}$. Note that the significance of $\tilde{P}_n$ approaching zero is that it indicates that with less energy consumption hybrid NOMA can almost surely achieve a higher rate than OMA, i.e., higher energy efficiency can be achieved by hybrid NOMA.
  \item It can be concluded that the conditions for $\tilde{P}_n$ approaches zero by applying HSIC are more relaxed
  compared to FSIC, which demonstrates the importance for applying HSIC. Extensive numerical results are provided to verify the developed analysis and also demonstrate the superior performance of the proposed HSIC aided hybrid NOMA scheme. The impact of user pairing is also discussed.
\end{itemize}

\section{System model}
Consider an uplink communication scenario with one BS and $M$ users, each denoted by $U_i$, $1\le i\le M$.
In the legacy OMA based network, TDMA is considered where each user is allocated with an individual time slot whose duration is $T$. Without loss of generality, in each frame, the $n$-th time slot is assigned to $U_n$.  Thus, in the legacy TDMA network, the amount of data that user $n$ can successfully upload to the BS is given by $\log(1+\rho_n|h_n|^2)$,
where $\rho_n$ is the transmit power of $U_n$, and $h_n$ is the normalized small scale Rayleigh fading gain.
{\color{black}Please note that the background noise is normalized to be $1$ in this paper for notational simplicity.}
Besides, it is noteworthy that slow time varying channel is considered and it is assumed that the
channel gain of a user in a frame remains a constant. Moreover, the users are ordered according to their channel
gains as follows\footnote{{\color{black}
The motivation for ordering users is to reveal the impact of user pairing on the performance of the proposed HSIC aided hybrid NOMA scheme.}}:
\begin{align}\label{eq1}
	{\color{black}\left | h_{1} \right | ^2
\leq \left | h_{2} \right | ^2
\leq\cdots \leq \left |  h_{M}\right|^2.}
\end{align}

Note that, in the legacy TDMA network, each user have only one chance to transmit data in a frame. In this paper,
Hybrid NOMA scheme is considered to provide more opportunities to transmit for users.
Note that the considered hybrid NOMA scheme can be implemented as a simple add-on to the legacy network. In the considered hybrid NOMA scheme, the $M$ users are classified into two types: the legacy users and the opportunistic users.
Each legacy user can only transmit data by using its own time slot as in TDMA. Whereas each opportunistic user can transmit data not only by using its own slot, but also sharing another legacy user's time slot by applying NOMA. Particularly, the opportunistic user $U_n$ is grouped with the legacy user $U_m$. Thus, $U_n$ can transmit signal
in the $m$-th time slot (say ``NOMA transmission''), as well as in the $n$-th time slot (say ``OMA transmission'').
To guarantee that the considered hybrid NOMA scheme not consumes more power than the benchmark TDMA scheme, $U_n$ can only transmit with power $\beta\rho_n$ in both the $m$-th and $n$-th slots, where $0<\beta<\frac{1}{2}$.

{\color{black}The rationality for the classification of legacy users and opportunistic users  originates from the diversity of service requirements of different kind of wireless users. The legacy user is more sensitive to transmission reliability instead of
the amount of data. For example, the legacy user could be a sensor, which needs to
upload fixed amount of data to the network periodically. In contrast, an opportunistic
user desires to transmit as much data as possible, which could possibly be an enhanced mobile broadband (eMBB) user such as virtual reality (VR) or augmented reality (AR) devices. Therefore, if the admission of the opportunistic user to the legacy user's time slot does not degrade the transmission reliability of the legacy user, it is reasonable
to offer the opportunistic user an additional transmission chance
by sharing the legacy user's time slot. Besides, for the NOMA transmission phase,
the legacy and opportunistic users can be regraded as ``primary user'' and ``secondary user'', respectively,
inspired by the concept of cognitve radio (CR) network.}\footnote{{\color{black}Note that 
this paper only focuses on the performance improvement for the opportunistic users by applying hybrid NOMA, 
where the service requirements of legacy and opportunistic users are different. 
It is also important to investigate whether hybrid NOMA can boost all users' performance, 
where all considered users in the network have symmetric service requirements.}}

\subsection{{\color{black}NOMA transmission in the $m$-th time slot}}
For the NOMA transmission in the $m$-th time slot, it is noteworthy that the transmission of $U_n$ should not
degrade the transmission reliability of $U_m$ compared to the TDMA scheme.
More specifically, there's a preset target rate denoted by $R_m$ which is the minimal rate
for ensuring the successful transmission of $U_m$. The transmission of $U_n$ should guarantee that
the outage probability of $U_m$ is the same as in OMA. To meet the aforementioned requirement,
the following will present two schemes,
one is the benchmark scheme considered in the pioneer work \cite{Ding2019MEC} named the FSIC scheme,
followed by the scheme proposed in this paper named the HSIC scheme.
{\color{black}Note that it is assumed that the channel state information (CSI) and transmit power of $U_m$
will be sent to $U_n$
prior to transmission. Besides, the opportunistic user $U_n$ can obtain its own CSI via the pilot
transmited by the BS.}

\subsubsection{{\color{black}FSIC Scheme for NOMA transmission}}
{\color{black}SIC is carried out at the BS, and the decoding of $U_n$'s signal is fixed at the first stage of SIC.} To ensure the transmission of $U_n$ is transparent to $U_m$,  the achievable rate of $U_n$ is capped to
$\bar{R}_{n,1}=\log(1+\frac{\beta \rho_n |h_n|^2}{\rho_m|h_m|^2+1})$. It can be easily seen that
if $U_n$ transmits with rate $\bar{R}_{n,1}$, then the success of SIC can be guaranteed and $U_m$'s signal can be decoded without any interference as in OMA.

Note that in FSIC scheme, $U_n$ is always put at the first stage of SIC. However, when the channel gain of $U_m$ is relatively large, the resulting achievable rate of $U_n$ could be very low. In this case, as an alterative, putting
$U_n$ at the second stage of SIC might yield higher rate. Motivated by this, it is potential to dynamically determine the decoding order of $U_n$ according to the channel conditions, as stated in the following HSIC scheme.
\subsubsection{{\color{black}HSIC Scheme for NOMA transmission}}
the decoding order of $U_n$ is determined according to the relationship between the received power of $U_n$ and
a threshold denoted by $\tau_m$, which is given by:
\begin{align}
\tau_m=\max\left \{ 0,\frac{\rho_{ m}|h_m|^2 }{2^{R_{ m}} -1} -1\right \}.
\end{align}
Note that $\tau_m$ can be interpreted as the maximum interference, with which $U_m$ can
still achieve the same outage performance as in OMA \cite{sun2021new,ding2021new}.
Based on the relationship of $\beta\rho_{n}\left | h_{n} \right |^2$ and $\tau_m$, the SIC strategies in the $m$-th slot can be classified into the following two types:
\begin{itemize}
  \item Type I: the received signal power of $U_n$ at the BS is less than or equal to $\tau(m)$, i.e.,
  $\beta\rho_{n}\left | h_{n} \right |^2 \le \tau_m $. For this case, putting $U_n$ at the second stage of SIC
  can yield larger rate compared to putting $U_n$ at the first stage of SIC, and will not hinder the user m from successfully decoding its signal. Thus, it is favorable to decode $U_n$'s signal at the second stage of SIC, and the achievable rate of $U_n$ is given by:
	$\tilde{R}_{n,1}^I=\log(1+\beta\rho_{ n}\left | h_{ n}  \right |^2)$.
  \item Type II: the received signal power of $U_n$ at the BS is larger than $\tau_m$, i.e., $\beta \rho_{n}\left | h_{n} \right |^2 > \tau_m$. For this case, to not degrade the QoS of $U_m$, $U_n$ can only be decoded at the first stage of SIC in HSIC, yielding the following achievable rate of $U_n$:
      $\tilde{R}_{n,1}^{II}=\log(1+\frac{\beta \rho_{n}\left | h_{n} \right |^2 }{\rho_{m}\left | h_m \right | ^2+1} )$.
Therefore, the achievable rate of $U_n$ during the time slot $T_m$ can be expressed as follows:
\begin{align}
\tilde{R}_{n,1}=\begin{cases}
		\log(1+\beta\rho_{ n}\left | h_{ n}  \right |^2) ,\!&\beta \rho_{n}\left | h_{n} \right |^2\! \le \! \tau_m \\
		\log(1+\frac{\beta \rho_{n}\left | h_{n} \right |^2 }{\rho_{m}\left | h_m \right | ^2+1} ),\!&\beta \rho_{n}\left | h_{n} \right |^2\! > \!\tau_m.
	\end{cases}
\end{align}
\end{itemize}

{\color{black}Please note that the advanatge of HSIC compared to
FSIC can be explained as follows. In FSIC, the opportunistic user's signal can be decoded
only at the first stage of SIC, to make the decoding of the legacy user's signal as same
as in OMA. However, when the channel gain of the legacy user is very large,
the opportunistic user will suffer from a strong interference from the legacy user, yielding
a very low data rate of the opportunistic user. However, for this case, if the opportunistic user
is decoded at the second stage of SIC and the interference caused by the opportunistic user's signal
is tolerable for the legacy user, i.e., $\rho_n|h_n|^2<\tau_m$, the opportunistic user can achieve
a larger a data rate. }

\subsection{\color{black}OMA transmission in the $n$-th time slot}
For the OMA transmission in the $n$-th time slot, $U_n$ exclusively occupies the time slot, resulting in the following achievable rate:
$R_{n,2}=\log(1+\beta \rho_n |h_n|^2)$.

{\color{black}From the above descriptions, it can be easily obtained that the achievable data rates of FSIC aided hybrid NOMA and HSIC aided hybrid NOMA can be expressed as $T\bar{R}_{n,1}+TR_{n,2}$ and $T\tilde{R}_{n,1}+T{R}_{n,2}$, respectively.}

Note that it has been shown that HSIC can help to improve transmission reliability for single time slot NOMA \cite{ding2021new}.
{\color{black}Differently and interestingly, this paper will demonstrate that HSIC can also be applied to hybrid NOMA for improving energy efficiency. Based on the above discussions, it can be easily found that the energy consumption for both the FSIC aided hybrid NOMA and HSIC aided NOMA is $2T\beta\rho_n$, which is lower than the OMA counterpart, since $0<\beta<\frac{1}{2}$. Thus, for a given $\beta$, if the achievable rate of hybrid NOMA is not lower than
that of OMA, then it can be concluded that hybrid NOMA is more energy efficient, since energy efficiency can be denoted by the ratio of data rate to
energy consumption.} Hence, it is interesting to
study the following probabilities that the achievable rate of hybrid NOMA is not larger than that of its OMA counterpart:
\begin{align}
\bar{P}_n=\text{Pr}\left( T\bar{R}_{n,1}+TR_{n,2}\le T \log(1+\rho_n|h_n|^2) \right),
\end{align}
for FSIC aided hybrid NOMA and
\begin{align}\label{Pn_H}
\tilde{P}_n=\text{Pr}\left( T\tilde{R}_{n,1}+T{R}_{n,2}\le T \log(1+\rho_n|h_n|^2) \right).
\end{align}
for HSIC aided hybrid NOMA.

\section{Performance analysis for HSIC aided hybrid NOMA scheme}
In this section, closed-form expression for $\tilde{P}_n$ is obtained, by considering both cases where $m>n$ and
$m<n$, based on which asymptotic analysis is further carried out in the high SNR regime. Moreover, interesting insights are obtained by comparing the asymptotic performance of HSIC aided hybrid NOMA with that of FSIC aided hybrid NOMA, which indicates the benefit of applying HSIC.

To obtain the expression for $\tilde{P}_n$, it is important to observe the following lemma, which divides
$\tilde{P}_n$ into three parts.

\begin{lemma}
  For ease of calculation, the probability $\tilde{P}_n$ can be expressed as the sum of the four following probabilities: $P_{1,1}$, $P_{1,2}$, $P_{2,1}$ and $P_{2,2}$. Define $P_1=P_{1,1}+P_{1,2}$. $\tilde{P}_n$ can be further simplified as follows:
	\begin{align}
		\tilde{P}_n=&\underbrace{P\left( |h_n|^2\le \Phi(|h_m|^2),\alpha_m<|h_m|^2< \frac{1-\beta}{\beta}\alpha_m\right)}_{P_{1,1}}\notag\\
&+\underbrace{P\left(  |h_n|^2\le \frac{1-2\beta}{\beta^2\rho_n}, |h_m|^2> \frac{1-\beta}{\beta}\alpha_m \right)}_{P_{1,2}}\notag\\
&+\underbrace{P\left( |h_n|^2\!\le\! \Psi(|h_m|^2),|h_n|^2\!>\!\Phi(|h_m|^2),|h_m|^2\!>\!\alpha_m \right)}_{P_{2,1}}\notag\\
&+\underbrace{P\left(  |h_n|^2\le \Psi(|h_m|^2), |h_m|^2<\alpha_m  \right)}_{P_{2,2}},
	\end{align}
where {\color{black}$\epsilon_m=2^{R_m}-1$, $\alpha_m=\frac{\epsilon_m}{\rho_m}$,} $\Phi(|h_m|^2)=\frac{|h_m|^2 \alpha_m^{-1}-1}{\beta\rho_n}$, and  $\Psi(|h_m|^2)=\frac{(1-\beta)(\rho_m |h_m|^2+1)-\beta}{\beta^2\rho_n}$.
\begin{IEEEproof}
	Please refer to Appendix A.	
\end{IEEEproof}
\end{lemma}

Based on Lemma $1$, the expressions for $\tilde{P}_n$ under the cases where $m<n$ and $m>n$ can be obtained, {\color{black}which are shown in the following two subsections, respectively.}

\subsection{Analytical results for $m< n$}
\begin{Theorem}
When $m<n$,  $\tilde{P}_n$  can be expressed as
\begin{align}
\tilde{P}_n=P_1+P_{2,1}+P_{2,2}{\color{black},}
\end{align}
where the expressions for $P_1$, $P_{2,1}$ and $P_{2,2}$ under the case where $\epsilon_m>\frac{\beta}{1-\beta}$ and the case where $\epsilon_m\le\frac{\beta}{1-\beta}$ are summarized by table I and table II, respectively. In Tables I and II, the expressions for the variables are given as follows:
\begin{align}
\kappa_1=\frac{1-2\beta}{(1-\beta)\beta\epsilon_m}{\color{black},}\\
\kappa_2=\frac{1-2\beta}{\beta^2\epsilon_m}+\frac{1-\beta}{\beta^2}{\color{black},}\\
\kappa_3=\frac{1-2\beta}{(1-\beta)\beta\epsilon_m}+\frac{1}{\beta}{\color{black}.}
\end{align}

\begin{table*}
\centering
\begin{threeparttable}
\label{tablename}
\caption{The expressions for $P_1$, $P_{2,1}$ and $P_{2,2}$, when $m<n$ and $\epsilon_m>\frac{\beta}{1-\beta}$}
\vspace{-1pt}
\renewcommand{\arraystretch}{1.5}
\begin{tabular}{l|c|c|c|c|c|c}
\hline
&
   &$\frac{\rho_n}{\rho_m}\le \kappa_1$ &$\kappa_1<\frac{\rho_n}{\rho_m}\le \frac{1}{\beta\epsilon_m}$ &$\frac{1}{\beta\epsilon_m}<\frac{\rho_n}{\rho_m}\le \frac{1-\beta}{\beta^2}$
   &$\frac{1-\beta}{\beta^2}<\frac{\rho_n}{\rho_m}\le \kappa_2 $
&$\frac{\rho_n}{\rho_m}> \kappa_2 $\\
\hline
\multirow{3}{*}{Exact results} &$P_{1}$ &$T_1$ &0 &0 &0 &0  \\
\cline{2-7}
&$P_{2,1}$ &$T_2$ &$T_2$ &$T_3$  &$T_4$ &0\\
\cline{2-7}
&$P_{2,2}$ &$T_6$ &$T_6$ &$T_6$ &$T_6$ &$T_7$\\
\hline
\multirow{3}{*}{Approximations} &$P_{1}$ &$\frac{1}{\rho_m^n}\tilde{T}_1$ &0 &0 &0 &0  \\
\cline{2-7}
&$P_{2,1}$ &{\color{black}$\tilde{T}_2$} &{\color{black}$\tilde{T}_2$} &{\color{black}$\tilde{T}_3$}  &$\frac{1}{\rho_m^n}\tilde{T}_4$ &0\\
\cline{2-7}
&$P_{2,2}$ &$\frac{1}{\rho_m^n}\tilde{T}_6$ &$\frac{1}{\rho_m^n}\tilde{T}_6$ &$\frac{1}{\rho_m^n}\tilde{T}_6$ &$\frac{1}{\rho_m^n}\tilde{T}_6$ &$\frac{1}{\rho_m^n}\tilde{T}_7$\\
\hline
\end{tabular}
\end{threeparttable}
\end{table*}

\begin{table*}
\centering
\begin{threeparttable}
\label{tablename}
\caption{The expressions for $P_1$, $P_{2,1}$ and $P_{2,2}$, when $m<n$ and $\epsilon_m\le\frac{\beta}{1-\beta}$.}
\vspace{-1pt}
\renewcommand{\arraystretch}{1.5}
\begin{tabular}{l|c|c|c|c|c}
\hline
&
   &$\frac{\rho_n}{\rho_m}\le \kappa_1$
   &$\kappa_1<\frac{\rho_n}{\rho_m}\le \kappa_3$ &$\kappa_3<\frac{\rho_n}{\rho_m}\le \kappa_2 $
&$\frac{\rho_n}{\rho_m}> \kappa_2 $\\
\hline
\multirow{3}{*}{\quad\quad Exact results} &$P_{1}$ &$T_1$ &0 &0 &0 \\
\cline{2-6}
&$P_{2,1}$ &$T_5$ &$T_5$ &$T_4$  &$0$\\
\cline{2-6}
&$P_{2,2}$ &$T_6$ &$T_6$ &$T_6$ &$T_7$\\
\hline
\multirow{3}{*}{\quad\quad Approximations}
&$P_{1}$ &$\frac{1}{\rho_m^n}\tilde{T}_1$ &0 &0 &0 \\
\cline{2-6}
&$P_{2,1}$ &$\frac{1}{\rho_m^n}\tilde{T}_5$ &$\frac{1}{\rho_m^n}\tilde{T}_5$ &$\frac{1}{\rho_m^n}\tilde{T}_4$  &0\\
\cline{2-6}
&$P_{2,2}$ &$\frac{1}{\rho_m^n}\tilde{T}_6$ &$\frac{1}{\rho_m^n}\tilde{T}_6$ &$\frac{1}{\rho_m^n}\tilde{T}_6$ &$\frac{1}{\rho_m^n}\tilde{T}_7$\\
\hline
\end{tabular}
\end{threeparttable}
\end{table*}

\begin{align}
T_1=&c_{mn}\!\sum\limits_{p=0}^{n\!-\!m\!-\!1}\!\!c_p\!\sum\limits_{l=0}^{m\!-\!1}\!c_l\frac{1}{M-m-p} \times\notag\\
&\Big(\!{\color{black}\varphi(\omega_1,\omega_3,M\!-\!m\!+\!l\!+\!1)}-e^{\!\frac{M-m-p}{\beta\rho_n}}\!{\color{black}\varphi(\omega_1,\omega_2,r)}\!\! \notag\\ & -e^{-(M-m-p)\omega_3}{\color{black}\varphi(\omega_2,\omega_3,l\!+\!p\!+\!1) } \!\Big),
\end{align}

 where $c_{mn}\!=\!\frac{M!}{(m-1)!(n-m-1)!(M-n)!}$, $c_l\!=\!\!\left(\!\!\begin{array}{c}
		m\!-\!1\\
		l
	\end{array}\!\!\right)\!(-1)^l$,
and $c_p\!=\!\left(\!\!\!\begin{array}{c}
		n\!-\!m\!-\!1\\
		l
	\end{array}\!\!\!\right)\!(-1)^{n-m-1-p}$, $\omega_1=\frac{1}{\alpha_m^{-1}-\beta\rho_n}$,
$\omega_2=\frac{1-\beta}{\beta}\alpha_m$, $\omega_3=\frac{1-2\beta}{\beta^2\rho_n}$, $r=l+p+1+\frac{M-m-p}{\beta\rho_n\alpha_m}$,
{\color{black}$\varphi(x,y,z) = \frac{e^{-xz}-e^{-yz}}{z}$,}
\begin{align}
T_2=&c_{mn}\! \sum\limits_{p=0}^{n\!-\!m\!-\!1}\!\!c_p\!\sum\limits_{l=0}^{m\!-\!1}\!c_l\frac{1}{M\!-\!m\!-\!p}
 \Big(\!{\color{black}\varphi(\alpha_m,\omega_1,M\!-\!m\!+\!l\!+\!1)}\notag \\ & \!-e^{-(M-m-p)\omega_3}\frac{e^{-a\alpha_m}}{a}+e^{\frac{M-m-p}{\beta\rho_n}}\frac{e^{-r\omega_1}}{r}  \!\Big),
\end{align}
where $a=l+p+1+\frac{(M-m-p)(1-\beta)\rho_m}{\beta^2\rho_n}$,
\begin{align}
T_3\!=&c_{mn}\!\!\!\sum\limits_{p=0}^{n\!-\!m\!-\!1}\!\!\!\!c_p\!\!\sum\limits_{l=0}^{m\!-\!1}\!c_l\frac{1}{M\!-\!m\!-\!p}
\times \notag\\
&
\Big(\frac{e^{-(M\!-\!m+l+1)\alpha_m}}{M\!-\!m+l+1}\!-\!e^{-(M\!-\!m\!-\!p)\omega_3}\!\frac{e^{\!-a\alpha_m}\!}{a} \Big),
\end{align}
\begin{align}
T_4\!=&c_{mn}\!\!\!\sum\limits_{p=0}^{n\!-\!m\!-\!1}\!\!\!\!c_p\!\!\sum\limits_{l=0}^{m\!-\!1}\!c_l\frac{1}{M\!-\!m\!-\!p}
\Big({\color{black}\varphi(\alpha_m,\omega_4,M\!-\!m\!+\!l\!+\!1)}\notag \\ &-\!e^{-(M\!-\!m\!-\!p)\omega_3}{\color{black}\varphi(\alpha_m,\omega_4,a) } \Big),
\end{align}
where $\omega_4=\frac{1-2\beta}{\beta^2\rho_n-(1-\beta)\rho_m}$,
\begin{align}
T_5\!=&c_{mn}\!\!\!\sum\limits_{p=0}^{n\!-\!m\!-\!1}\!\!\!\!c_p\!\!\sum\limits_{l=0}^{m\!-\!1}\!c_l\frac{1}{M\!-\!m\!-\!p}
\Big(\!{\color{black}\varphi(\alpha_m,\omega_1,M\!-\!m\!+\!l\!+\!1)}\notag \\ &- e^{-(M\!-\!m\!-\!p)\omega_3}{\color{black}\varphi(\alpha_m,\omega_5,a)} \!  +\!e^{\frac{M\!-\!m\!-\!p}{\beta\rho_n}} {\color{black}\varphi(\omega_1,\omega_5,r)}
\Big),
\end{align}
where $\omega_5=\frac{1-\beta}{\beta\alpha_m^{-1}-(1-\beta)\rho_m}$.
\begin{align}
T_6\!=&c_{mn}\!\!\!\sum\limits_{p=0}^{n\!-\!m\!-\!1}\!\!\!c_p\!\sum\limits_{l=0}^{m\!-\!1}\!c_l\frac{1}{M\!-\!m\!-\!p}\times \\
&\Big(\!{\color{black}\varphi(0,\alpha_m,M\!-\!m\!+\!l\!+\!1)}\!-\!e^{-(M\!-\!m\!-\!p)\omega_3}{\color{black}\varphi(0,\alpha_m,a)
}\Big)\!,\notag
\end{align}
\begin{align}
T_7\!=&c_{mn}\!\!\!\sum\limits_{p=0}^{n\!-\!m\!-\!1}\!\!\!c_p\!\sum\limits_{l=0}^{m\!-\!1}\!c_l\frac{1}{M\!-\!m\!-\!p}\times \\
&\Big(\!{\color{black}\varphi(0,\omega_4,M\!-\!m\!+\!l\!+\!1)}\!-\!e^{-(M\!-\!m\!-\!p)\omega_3}\!{\color{black}\varphi(0,\omega_4,a)}
\Big).\notag
\end{align}
\begin{IEEEproof}
	Please refer to Appendix B.	
\end{IEEEproof}
\end{Theorem}

By applying Taylor expansions, the extreme value of $\tilde{P}_n$ in the high SNR regime can be obtained, as highlighted in the following corollary.

\begin{Corollary}
For the case where $m<n$, when $\rho_n \to \infty$, $\rho_m \to \infty$, and $\frac{\rho_n}{\rho_m}=\eta$ is a constant, the value of $\tilde{P}_n$ for $\epsilon_m>\frac{\beta}{1-\beta}$ and $\epsilon_m
\leq \frac{\beta}{1-\beta}$ can be approximated as summarized in table I  and table II, respectively.
The expressions for the variables used in Table I and Table II can be expressed as follows:
\begin{align}
\tilde{T}_1\! = &c_{mn}\!\!\!\sum\limits_{p=0}^{n\!-\!m\!-\!1}\!\!\!
\left(\!\!\begin{array}{c}
		n\!-\!m\!-\!1\\
		p
	\end{array}\!\!\right)\!\frac{(-1)^p}{m+p}
\Big(  \frac{\varpi_3^n-\varpi_1^n}{n} -\epsilon_m^{m+p}
\notag \\ &
\times
 \sum\limits_{q=0}^{m+p}\!\!\!
\left(\!\!\begin{array}{c}
		m+p\\
		q
	\end{array}\!\!\right)\!(\beta\eta)^{m+p-q}\frac{\varpi_3^{n-q}-\varpi_1^{n-q}}{n-q}  \Big),
\end{align}
where $\varpi_1=\frac{1}{\epsilon_m^{-1}-\beta\eta}$, and $\varpi_3=\frac{1-2\beta}{\beta^2\eta}$, and note that $\varpi_1$ and $\varpi_3$ is not a function of $\rho_m$ or $\rho_n$.
\begin{align}
\tilde{T}_2=&c_{mn}\!\sum\limits_{p=0}^{n\!-\!m\!-\!1}\!\!c_p\!\sum\limits_{l=0}^{m\!-\!1}\!c_l\frac{1}{M\!-\!m\!-\!p}
\Big(\frac{1}{r}-\frac{1}{a}\Big),
\end{align}
\begin{align}
\tilde{T}_3=&c_{mn}\!\!\!\sum\limits_{p=0}^{n\!-\!m\!-\!1}\!\!c_p\!\sum\limits_{l=0}^{m\!-\!1}\!c_l\frac{1}{M\!-\!m\!-\!p}
\Big(\frac{1}{M-m+l+1}-\frac{1}{a}\Big),
\end{align}
\begin{align}
\tilde{T}_4\! = &c_{mn}\!\!\!\sum\limits_{p=0}^{n\!-\!m\!-\!1}\!\!\!
\left(\!\!\begin{array}{c}
		n\!-\!m\!-\!1\\
		p
	\end{array}\!\!\right)\!\frac{(-1)^p}{m+p}
\Big(  \frac{\varpi_4^n-\epsilon_m^n}{n} \notag \\ &-\epsilon_m^{m+p}\frac{\varpi_6^{n-m-p}-\epsilon_m^{n-m-p}}{n-m-p}-\sum\limits_{q=0}^{m+p}\!\!\!
\left(\!\!\begin{array}{c}
		m+p\\
		q
	\end{array}\!\!\right)\! \notag \\ & \frac{(-1)^q(1\!-\!2\beta)^q(\beta^2\eta)^{m+p-q}(\varpi_4^{n\!-\!q}\!\!-\!\varpi_6^{n\!-\!q})}{(1-\beta)^{m+p}(n-q)}   \Big),
\end{align}
where $\varpi_4=\frac{1-2\beta}{\beta^2\eta-(1-\beta)}$.
\begin{align}
\tilde{T_5}\! = &c_{mn}\!\!\!\sum\limits_{p=0}^{n\!-\!m\!-\!1}\!\!\!
\left(\!\!\begin{array}{c}
		n\!-\!m\!-\!1\\
		p
	\end{array}\!\!\right)\!\frac{(-1)^p}{m+p}
\Big(  \frac{\varpi_1^n-\epsilon_m^n}{n} \notag \\ & -\epsilon_m^{m+p}\frac{\varpi_6^{n-m-p}-\epsilon_m^{n-m-p}}{n-m-p}+\sum\limits_{q=0}^{m+p}\!\!\!
\left(\!\!\begin{array}{c}
		m+p\\
		q
	\end{array}\!\!\right)\! \notag \\ & \Big(     \epsilon_m^{m+p}(\beta\eta)^{m+p-q}\frac{\varpi_7^{n-q}-\varpi_1^{n-q}}{n-q}     \notag \\ & -  \frac{(-1)^q(1\!-\!2\beta)^q(\beta^2\eta)^{m+p-q}(\varpi_7^{n\!-\!q}\!\!-\!\varpi_6^{n\!-\!q})}{(1-\beta)^{m+p}(n-q)}        \Big)\Big),
\end{align}
where $\varpi_6=\frac{(1-\beta)\epsilon_m+1-2\beta}{\beta^2\eta}$,
and $\varpi_7=\frac{\varpi_5\epsilon_m^{-1}-1}{\beta\eta}$.
\begin{align}
\tilde{T}_6\! = &c_{mn}\!\!\!\sum\limits_{p=0}^{n\!-\!m\!-\!1}\!\!\!
\left(\!\!\begin{array}{c}
		n\!-\!m\!-\!1\\
		p
	\end{array}\!\!\right)\!\frac{(-1)^p}{m+p}\notag \\ &
\times\Big(  \frac{\epsilon_m^n}{n}
+ \epsilon_m^{m+p} \frac{\varpi_6^{n-m-p}-\epsilon_m^{n-m-p}}{n-m-p} -\sum\limits_{q=0}^{m+p}\!\!\!
\left(\!\!\begin{array}{c}
		\!m+p\!\\
		q
	\end{array}\!\!\right)\! \notag \\ & \times \frac{(-1)^q(1\!-\!2\beta)^q(\beta^2\eta)^{m+p-q}(\varpi_6^{n\!-\!q}\!\!-\!\varpi_3^{n\!-\!q})}{(1-\beta)^{m+p}(n-q)}  \Big),
\end{align}
\begin{align}
\tilde{T_7}\! =&c_{mn}\!\!\!\sum\limits_{p=0}^{n\!-\!m\!-\!1}\!\!\!
\left(\!\!\begin{array}{c}
		n\!-\!m\!-\!1\\
		p
	\end{array}\!\!\right)\!\frac{(-1)^p}{m+p} \Big( \frac{\varpi_4^n}{n}
- \sum\limits_{q=0}^{m+p}\!\!\!
\left(\!\!\begin{array}{c}
		\!\!m+p\!\!\\
		q
	\end{array}\!\!\right)\! \notag \\ & \times \frac{(-1)^q(1\!-\!2\beta)^q(\beta^2\eta)^{m+p-q}(\varpi_4^{n\!-\!q}\!\!-\!\varpi_3^{n\!-\!q})}{(1-\beta)^{m+p}(n-q)}   \Big).
\end{align}
\begin{IEEEproof}
	Please refer to Appendix C.	
\end{IEEEproof}
\end{Corollary}

Based on the results shown in Corollary $1$, interesting insights can be obtained as highlighted in the
following remarks.
\begin{Remark}
From the results shown in Tables I and II, it can be easily observed that except the case
where $\epsilon_m>\frac{\beta}{1-\beta}$ and $\frac{\rho_n}{\rho_m}\le \frac{1-\beta}{\beta^2}$ both hold,
$\tilde{P}_n\to 0$ when $\rho_n$ and $\rho_m$ goes infinity. By contrast, in the FSIC aided hybrid NOMA scheme, $\bar{P}_n$ can approach zero in the high SNR regime only when $\frac{\rho_n}{\rho_m}>\frac{1-\beta}{\beta^2}$\cite{Ding2019MEC}. Thus, the conditions for $\tilde{P}_n$ of HSIC aided hybrid NOMA to approach zero is relaxed compared to FSIC.
\end{Remark}

{\color{black}Note that the meaning for the probability $\tilde{P}_n$ can approache zero is that the probability that
hybrid NOMA can achieve a higher rate than OMA can approach one with less energy consumption (for any given $\beta<1$),
which indicates the superior potential of hybrid NOMA in terms of energy efficiency improvement.}

\begin{Remark}
For the case where $\rho_n \to \infty$, $\rho_m \to \infty$, and $\frac{\rho_n}{\rho_m}$ is a constant,
when $\tilde{P}_n$ goes zero, i.e., when $\epsilon_m<\frac{\beta}{1-\beta}$ or when $\epsilon_m>\frac{\beta}{1-\beta}$ and $\frac{\rho_n}{\rho_m}> \frac{1-\beta}{\beta^2}$,
it can be observed that $\tilde{P}_n$ goes zero with exponential decaying rate $n$, i.e.,
\begin{align}
\tilde{P}_n\propto \frac{1}{\rho_m^n}, \text{ or }\tilde{P}_n\propto \frac{1}{\rho_n^n},
\end{align}
which means that the impact of $n$ (i.e., the order of the channel gain of the opportunistic user) on $\tilde{P}_n$ is dominated comparing to $m$.
\end{Remark}

It is also interesting to investigate the behavior of $\tilde{P}_n$ in another two limiting cases,
i.e., when  $\rho_m\to\infty$ while $\rho_n$ is a constant and when $\rho_n\to\infty$ while $\rho_m$ is a constant.
When  $\rho_m\to\infty$ while $\rho_n$ is a constant, by noting that $\frac{\rho_n}{\rho_m}<\kappa_1$ and applying Taylor series, the following corollary is not hard to be obtained.

\begin{Corollary}
When $\rho_m\to\infty$ and $\rho_n$ is a constant, $\tilde{P}_n$ for $m<n$ approaches a non-zero constant as given by:
\begin{align}
\tilde{P}_n\!\approx &c_{mn}\!\!\!\sum\limits_{p=0}^{n\!-\!m\!-\!1}\!\!\!\!c_p\!\!\sum\limits_{l=0}^{m\!-\!1}\!c_l\frac{1}{M\!-\!m\!-\!p}
\Big({\color{black}\varphi(0,\omega_3,M\!-\!m\!+\!l\!+\!1)} \\
& -e^{-(M-m-p)\frac{1-2\beta}{\beta^2\rho_n}}{\color{black}\varphi(0,\omega_3, l\!+\!p\!+\!1)}\Big).\notag
\end{align} 
\end{Corollary}

When $\rho_n\to\infty$ and $\rho_m$ is a constant, by noting that $\frac{\rho_n}{\rho_m} > \kappa_2$
and following the similar methods used in Appendix C, the following corollary can be obtained.

\begin{Corollary}
When $\rho_n\to\infty$ and $\rho_m$ is a constant,
$\tilde{P}_n$ for $m<n$ can be approximated as follows:
\begin{align}\label{Ext_P_n_inf_fin}
\tilde{P}_n\!\approx &\frac{1}{\rho_n^n}
\frac{c_{mn}\!\!\!\sum\limits_{p=0}^{n\!-\!m\!-\!1}\!\!\!\left(\!\!\!\begin{array}{c}
		n\!-\!m\!-\!1\\
		p
	\end{array}\!\!\!\right)\!\!(-1)^p}{n\!-\!m\!-\!p}\!\!\left(\!\frac{1}{m\!+\!p}\!-\!\frac{1}{n}\!\right)\!\!\frac{(1\!-\!2\beta)^n}{\beta^{2n}}.
\end{align}
\end{Corollary}

It can be straightforwardly seen from (\ref{Ext_P_n_inf_fin}) that $\tilde{P}_n$ approaches zero with a exponential decaying rate $n$, when $\rho_n\to\infty$ and $\rho_m$ is a constant.

\subsection{Analytical results $m>n$}
\begin{Theorem}
When {\color{black}$m>n$},  $\tilde{P}_n$  can be expressed as
\begin{align}
\tilde{P}_n=P_1+P_{2,1}+P_{2,2},
\end{align}
where the expressions for $P_1$, $P_{2,1}$ and $P_{2,2}$ under the case where $\epsilon_m>\frac{\beta}{1-\beta}$ and the case where $\epsilon_m\le\frac{\beta}{1-\beta}$ are given by table III and table IV, respectively. In Tables III and IV, the expressions for the variables are given as follows:
\begin{table*}
\centering
\begin{threeparttable}
\label{tablename}
\caption{The expressions for $P_1$, $P_{2,1}$ and $P_{2,2}$, when $m>n$ and $\epsilon_m>\frac{\beta}{1-\beta}$.}
\vspace{-1pt}
\renewcommand{\arraystretch}{1.5}
\begin{tabular}{l|c|c|c|c|c|c}
\hline
&
   &$\frac{\rho_n}{\rho_m}\le \kappa_1$ &$\kappa_1<\frac{\rho_n}{\rho_m}\le \frac{1}{\beta\epsilon_m}$ &$\frac{1}{\beta\epsilon_m}<\frac{\rho_n}{\rho_m}\le \frac{1-\beta}{\beta^2}$
   &$\frac{1-\beta}{\beta^2}<\frac{\rho_n}{\rho_m}\le \kappa_2 $
&$\frac{\rho_n}{\rho_m}> \kappa_2 $\\
\hline
\multirow{3}{*}{Exact results} &$P_{1}$ &$Q_1$ &$Q_2$ &$Q_2$ &$Q_2$ &$Q_2$  \\
\cline{2-7}
&$P_{2,1}$ &$Q_3$ &$Q_3$ &$Q_4$  &$Q_5$ &$Q_6$\\
\cline{2-7}
&$P_{2,2}$ &$Q_9$ &$Q_9$ &$Q_9$ &$Q_9$ &$Q_{10}$\\
\hline
\multirow{3}{*}{Approximations} &$P_{1}$ &$\frac{1}{\rho_m^{ n}}\tilde{S}\!+\!\frac{1}{\rho_m^m}\tilde{Q}_1$ &$\frac{1}{\rho_m^{ n}}\tilde{S}\!+\!\frac{1}{\rho_m^m}\tilde{Q}_2$ &$\frac{1}{\rho_m^{ n}}\tilde{S}\!+\!\frac{1}{\rho_m^m}\tilde{Q}_2$ &$\frac{1}{\rho_m^{ n}}\tilde{S}\!+\!\frac{1}{\rho_m^m}\tilde{Q}_2$ &$\frac{1}{\rho_m^{ n}}\tilde{S}\!+\!\frac{1}{\rho_m^m}\tilde{Q}_2$  \\
\cline{2-7}
&$P_{2,1}$ &$\frac{1}{\rho_m^m}\tilde{Q}_3$ &$\frac{1}{\rho_m^m}\tilde{Q}_3$  &{\color{black}$\tilde{Q}_4$} &{\color{black}$\tilde{Q}_5$} &{\color{black}$\tilde{Q}_6$}\\
\cline{2-7}
&$P_{2,2}$ &$\frac{1}{\rho_m^m}\tilde{Q}_9$ &$\frac{1}{\rho_m^m}\tilde{Q}_9$ &$\frac{1}{\rho_m^m}\tilde{Q}_9$ &$\frac{1}{\rho_m^m}\tilde{Q}_9$ &$\frac{1}{\rho_m^m}\tilde{Q}_{10}$\\
\hline
\end{tabular}
\end{threeparttable}
\end{table*}

\begin{table*}
\centering
\begin{threeparttable}
\label{tablename}
\caption{The expressions for $P_1$, $P_{2,1}$ and $P_{2,2}$, when $m>n$ and $\epsilon_m\le\frac{\beta}{1-\beta}$.}
\vspace{-1pt}
\renewcommand{\arraystretch}{1.5}
\begin{tabular}{l|c|c|c|c|c}
\hline
&
   &$\frac{\rho_n}{\rho_m}\le \kappa_1$
   &$\kappa_1<\frac{\rho_n}{\rho_m}\le \kappa_3$ &$\kappa_3<\frac{\rho_n}{\rho_m}\le \kappa_2 $
&$\frac{\rho_n}{\rho_m}> \kappa_2 $\\
\hline
\multirow{3}{*}{Exact results} &$P_{1}$ &$Q_1$ &$Q_2$ &$Q_2$ &$Q_2$ \\
\cline{2-6}
&$P_{2,1}$ &$Q_3$ &$Q_3$ &$Q_7$  &$Q_8$\\
\cline{2-6}
&$P_{2,2}$ &$Q_9$ &$Q_9$ &$Q_9$ &$Q_{10}$\\
\hline
\multirow{3}{*}{Approximations}
&$P_{1}$
&$\frac{1}{\rho_m^{ n}}\tilde{S}\!+\!\frac{1}{\rho_m^m}\tilde{Q}_1$
&$\frac{1}{\rho_m^{ n}}\tilde{S}\!+\!\frac{1}{\rho_m^m}\tilde{Q}_2$
&$\frac{1}{\rho_m^{ n}}\tilde{S}\!+\!\frac{1}{\rho_m^m}\tilde{Q}_2$
&$\frac{1}{\rho_m^{ n}}\tilde{S}\!+\!\frac{1}{\rho_m^m}\tilde{Q}_2$ \\
\cline{2-6}
&$P_{2,1}$ &$\frac{1}{\rho_m^m}\tilde{Q}_3$ &$\frac{1}{\rho_m^m}\tilde{Q}_3$ &$\frac{1}{\rho_m^m}\tilde{Q}_7$  &$\frac{1}{\rho_m^m}\tilde{Q}_8$\\
\cline{2-6}
&$P_{2,2}$ &$\frac{1}{\rho_m^m}\tilde{Q}_9$ &$\frac{1}{\rho_m^m}\tilde{Q}_9$ &$\frac{1}{\rho_m^m}\tilde{Q}_9$ &$\frac{1}{\rho_m^m}\tilde{Q}_{10}$\\
\hline
\end{tabular}
\end{threeparttable}
\end{table*}
 \begin{align}
Q_1=&\hat{c}_{mn}\!\!\!\sum\limits_{p=0}^{m\!-\!n\!-\!1}\!\!\hat{c}_p\!\sum\limits_{l=0}^{n\!-\!1}\!\hat{c}_l\frac{1}{l+p+1}
\Big(
{\color{black}\varphi(\alpha_m,\omega_3,M\!-\!n\!-\!p)}\\ & \!-e^{\!\frac{l\!+\!p\!+\!1}{\beta\!\rho_n}}\!
{\color{black}\varphi(\alpha_m,\omega_1,t)}\!\! -
{\color{black}\varphi(\omega_1,\omega_3,M\!-\!n+l+1)}\Big),\notag
\end{align}
where $\hat{c}_{mn}\!=\!\frac{M!}{(n-1)!(m-n-1)!(M-m)!}$,
 $\hat{c}_l\!=\!\!\left(\!\!\begin{array}{c}
		m\!-\!1\\
		l
	\end{array}\!\!\right)\!(-1)^l$,
 $\hat{c}_p\!=\!\left(\!\!\!\begin{array}{c}
		m\!-\!n\!-\!1\\
		l
	\end{array}\!\!\!\right)\!(-1)^{m-n-1-p}$, and $t=M-n-p+\frac{l+p+1}{\beta\rho_n\alpha_m}$.
\begin{align}
Q_2=&\hat{c}_{mn}\!\!\!\sum\limits_{p=0}^{m\!-\!n\!-\!1}\!\!\hat{c}_p\!\sum\limits_{l=0}^{n\!-\!1}\!\hat{c}_l\frac{1}{l+p+1}
\Big(
-e^{\frac{l+p+1}{\beta\rho_n}}
{\color{black}\varphi(\alpha_m,\omega_2,t)}\notag\\ & +\frac{e^{-(M\!-\!n\!-\!p)\alpha_m} \!\!-\! e^{-(l+p+1)\omega_3} e^{-(M\!-\!n\!-\!p)\omega_2}}{M-n-p}\Big),
\end{align}
\begin{align}
Q_3=&\hat{c}_{mn}\!\!\!\sum\limits_{p=0}^{m\!-\!n\!-\!1}\!\!\hat{c}_p\!\sum\limits_{l=0}^{n\!-\!1}\!\hat{c}_l\frac{1}{l+p+1}
\Big(
e^{\frac{l+p+1}{\beta\rho_n}}{\color{black}\varphi(\alpha_m,\omega_1,t)}\notag\\ & -{\color{black}\varphi(\alpha_m,\omega_1,M\!-\!n+l+1)}\Big),
\end{align}
\begin{align}
Q_4=&\hat{c}_{mn}\!\!\!\sum\limits_{p=0}^{m\!-\!n\!-\!1}\!\!\!\!\hat{c}_p\!\!\sum\limits_{l=0}^{n\!-\!1}\!\hat{c}_l\frac{1}{l+p+1}
\Big(
e^{\frac{l+p+1}{\beta\rho_n}}\!\frac{e^{-t\alpha_m}}{t} \notag\\
&\!-\!\frac{e^{-(M\!-\!n\!+l+1)\alpha_m} }{M\!-\!n\!+l+1}\Big),
\end{align}
\begin{align}
Q_5=&\hat{c}_{mn}\!\!\!\sum\limits_{p=0}^{m\!-\!n\!-\!1}\!\!\!\!\hat{c}_p\!\!\sum\limits_{l=0}^{n\!-\!1}\!\hat{c}_l\frac{1}{l+p+1}
\Big(
e^{\frac{l+p+1}{\beta\rho_n}}\!\frac{e^{-t\alpha_m}}{t} \!
\\ &
-\! e^{-(l+p+1)\omega_3}\frac{e^{-s\omega_4}}{s}
- {\color{black}\varphi(\alpha_m,\omega_4,M\!-\!n+l+1)}\Big), \notag
\end{align}
where $s=M-n-p+\frac{(l+p+1)(1-\beta)\rho_m}{\beta^2\rho_n}$,
\begin{align}
Q_6=&\hat{c}_{mn}\!\!\!\sum\limits_{p=0}^{m\!-\!n\!-\!1}\!\!\!\!\hat{c}_p\!\!\sum\limits_{l=0}^{n\!-\!1}\!\hat{c}_l\frac{1}{l+p+1}
\Big(
e^{\frac{l+p+1}{\beta\rho_n}}\!\frac{e^{\!-t\alpha_m\!}\!}{t} \!
\\
&-\! e^{\!-(l+p+1)\omega_3}\frac{\!e^{\!-s\alpha_m}}{s}\!\Big),\notag
\end{align}
\begin{align}
Q_7=&\hat{c}_{mn}\!\!\!\sum\limits_{p=0}^{m\!-\!n\!-\!1}\!\!\!\!\hat{c}_p\!\!\sum\limits_{l=0}^{n\!-\!1}\!\hat{c}_l\frac{1}{l+p+1}
\Big(
e^{\frac{l+p+1}{\beta\rho_n}}
{\color{black}\varphi(\alpha_m,\omega_5,t)}
\\
&-
{\color{black}\varphi(\alpha_m,\omega_4,M\!-\!n\!+\!l\!+\!1)\!}-\!e^{-(l+p+1)\omega_3}
{\color{black}\varphi(\omega_4,\omega_5,s)}
\Big),\notag
\end{align}
\begin{align}
Q_8=&\hat{c}_{mn}\!\!\!\sum\limits_{p=0}^{m\!-\!n\!-\!1}\!\!\!\!\hat{c}_p\!\!\sum\limits_{l=0}^{n\!-\!1}\!\hat{c}_l\frac{1}{l+p+1}
\Big(
e^{\frac{l+p+1}{\beta\rho_n}}
{\color{black}\varphi(\alpha_m,\omega_5,t)}
\notag \\
&- e^{-(l+p+1)\omega_3}
{\color{black}\varphi(\alpha_m,\omega_5,s)}
\Big),
\end{align} 
\begin{align}
Q_9\!=&\hat{c}_{mn}\!\!\!\sum\limits_{p=0}^{m\!-\!n\!-\!1}\!\!\!\!\hat{c}_p\!\!\sum\limits_{l=0}^{n\!-\!1}\!\hat{c}_l\frac{1}{l+p+1}
\Big(
{\color{black}\varphi(0,\alpha_m,M\!-\!n\!-\!p) \!}
\notag\\
& -{\color{black}\varphi(0,\alpha_m,M\!-\!n+l+1)}
\Big),
\end{align}
 \begin{align}
Q_{10}=&\hat{c}_{mn}\!\!\!\sum\limits_{p=0}^{m\!-\!n\!-\!1}\!\!\!\!\hat{c}_p\!\!\sum\limits_{l=0}^{n\!-\!1}\!\hat{c}_l\frac{1}{l+p+1}
\Big({\color{black}\varphi(0,\alpha_m,M\!-\!n\!-\!p)}
\\
&- {\color{black}\varphi(0,\omega_4,M\!-\!n+l+1)\!} - \! e^{(l+p+1)\omega_3}{\color{black}\varphi(\omega_4,\alpha_m,s)}
\Big).\notag
\end{align}
\end{Theorem}

\begin{IEEEproof}
	Please refer to Appendix D.	
\end{IEEEproof}

By applying Taylor expansions, the extreme value of $\tilde{P}_n$ in the high SNR regime can be obtained, as highlighted in the following corollary.

\begin{Corollary}
For the case where $m>n$, when $\rho_n \to \infty$, $\rho_m \to \infty$, and $\frac{\rho_n}{\rho_m}=\eta$ is a constant, the approximations for $\tilde{P}_n$ under the cases where $\epsilon_m>\frac{\beta}{1-\beta}$ and $\epsilon_m\leq\frac{\beta}{1-\beta}$ can be summarized as shown in Table III and Table IV, respectively. The variables used in Tables III and IV can be expressed as follows:
\begin{align}
\tilde{S} =\frac{\hat{c}_n\varpi_3^n}{n},
\end{align}
where $\hat{c}_n=\frac{M!}{(n-1)!(M-n)!}$.
\begin{align}
\tilde{Q}_1\!& =
\hat {c}_{mn}\!\!\!\sum\limits_{p=0}^{m\!-\!n\!-\!1}\!\!\!
\left(\!\!\begin{array}{c}
		\!m\!-\!n\!-\!1\!\\
		p
	\end{array}\!\!\right)\!\frac{(-1)^p}{n+p}\!
\Big(\frac{\varpi_3^m\!-\!\varpi_1^m}{m}\\
- &
\!\frac{\varpi_3^m}{m\!-\!n\!-\!p}
\! +\!
 \sum\limits_{q=0}^{n+p}\!\!
\left(\!\!\begin{array}{c}
		\!n\!+\!p\!\\
		q
	\end{array}\!\!\right)\!(-1)^q\frac{\varpi_1^{m-q}\!-\!\epsilon_m^{m-q}}{(\beta\eta)^{n+p}\epsilon_m^{n+p-q}(m-q)}  \Big).\notag
\end{align}
\begin{align}
\tilde{Q}_2 =&
\hat {c}_{mn}\!\!\!\sum\limits_{p=0}^{m\!-\!n\!-\!1}\!\!\!
\left(\!\!\begin{array}{c}
		\!m\!-\!n\!-\!1\!\\
		p
	\end{array}\!\!\right)\!\frac{(-1)^p}{n+p}
\Big( \sum\limits_{q=0}^{n+p}\!\!
\left(\!\!\begin{array}{c}
		\!n\!+\!p\!\\
		q
	\end{array}\!\!\right)\!(-1)^q\notag\\
& \frac{\varpi_2^{m-q}\!-\!\epsilon_m^{m-q}}{(\beta\eta)^{n+p}\epsilon_m^{n+p-q}(m-q)} -\varpi_3^{n+p}\frac{\varpi_2^{m-n-p}}{m-n-p} \Big).
\end{align}
\begin{align}
\tilde{Q}_3 =&
\hat {c}_{mn}\!\!\!\sum\limits_{p=0}^{m\!-\!n\!-\!1}\!\!\!
\left(\!\!\begin{array}{c}
		\!m\!-\!n\!-\!1\!\\
		p
	\end{array}\!\!\right)\!\frac{(-1)^p}{n+p}
\Big(\frac{\varpi_1^m-\epsilon_m^m}{m}- \notag\\
& \sum\limits_{q=0}^{n+p}\!\!
\left(\!\!\begin{array}{c}
		\!n\!+\!p\!\\
		q
	\end{array}\!\!\right)\!(-1)^q\frac{\varpi_1^{m-q}\!-\!\epsilon_m^{m-q}}{(\beta\eta)^{n+p}\epsilon_m^{n+p-q}(m-q)} \Big).
\end{align}
\begin{align}
\tilde{Q}_4=&\hat{c}_{mn}\!\sum\limits_{p=0}^{m\!-\!n\!-\!1}\!\!\!\!\hat{c}_p\!\!
\sum\limits_{l=0}^{n\!-\!1}\!\hat{c}_l\frac{1}{l+p+1}\!\Big(
\frac{1}{t}-\frac{1}{M-n+l+1}\Big),
\end{align}
\begin{align}
\tilde{Q}_5=&\hat{c}_{mn}\!\sum\limits_{p=0}^{m\!-\!n\!-\!1}\!\!\!\!\hat{c}_p\!\!
\sum\limits_{l=0}^{n\!-\!1}\!\hat{c}_l\frac{1}{l+p+1}\Big(
\frac{1}{t}-\frac{1}{s}\Big),
\end{align}
\begin{align}
\tilde{Q}_6=&\hat{c}_{mn}\!\sum\limits_{p=0}^{m\!-\!n\!-\!1}\!\!\!\!\hat{c}_p\!\!
\sum\limits_{l=0}^{n\!-\!1}\!\hat{c}_l\frac{1}{l+p+1}
\Big(\frac{1}{t}-\frac{1}{s}\Big),
\end{align}
\begin{align}
\tilde{Q}_7 =&
\hat {c}_{mn}\!\!\!\sum\limits_{p=0}^{m\!-\!n\!-\!1}\!\!\!
\left(\!\!\begin{array}{c}
		\!m\!-\!n\!-\!1\!\\
		p
	\end{array}\!\!\right)\!\frac{(-1)^p}{n+p}
\Big(\frac{\varpi_4^m-\epsilon_m^m}{m} \\ &
+\sum\limits_{q=0}^{n+p}\!\!
\left(\!\!\begin{array}{c}
		\!n\!+\!p\!\\
		q \end{array}\!\!\right)\!\Big(\frac{(1\!-\!2\beta)^q(1\!-\!\beta)^{n+p-q} (\varpi_5^{m-q}\!\!-\!\varpi_4^{m-q})}{(\beta^2\eta)^{n+p}(m-q)}\notag\\ &
-(-1)^q\frac{\varpi_5^{m-q}\!-\!\epsilon_m^{m-q}}{(\beta\eta)^{n+p}\epsilon_m^{n+p-q}(m-q)} \Big) \Big).\notag
\end{align}
\begin{align}
\tilde{Q}_8 =&
\hat {c}_{mn}\!\!\!\sum\limits_{p=0}^{m\!-\!n\!-\!1}\!\!\!
\left(\!\!\begin{array}{c}
		\!m\!-\!n\!-\!1\!\\
		p
	\end{array}\!\!\right)\!\frac{(-1)^p}{n+p}\times
\notag\\
&\sum\limits_{q=0}^{n+p}\!\!
\left(\!\!\begin{array}{c}
		\!n\!+\!p\!\\
		q \end{array}\!\!\right)\Big(\frac{(1\!-\!2\beta)^q(1\!-\!\beta)^{n+p-q} (\varpi_5^{m-q}\!\!-\!{\color{black}\epsilon_m^{m-q}})}{(\beta^2\eta)^{n+p}(m-q)}
\notag\\
& -(-1)^q\frac{\varpi_5^{m-q}\!-\!\epsilon_m^{m-q}}{(\beta\eta)^{n+p}\epsilon_m^{n+p-q}(m-q)} \Big).
\end{align} 
\begin{align}
\tilde{Q}_9 =&
\hat {c}_{mn}\!\!\!\sum\limits_{p=0}^{m\!-\!n\!-\!1}\!\!\!
\left(\!\!\begin{array}{c}
		\!m\!-\!n\!-\!1\!\\
		p
	\end{array}\!\!\right)\!\frac{(-1)^p}{n+p} \frac{\epsilon_m^m}{m}.
\end{align}
\begin{align}
\tilde{Q}_{10} =&
\hat {c}_{mn}\!\!\!\sum\limits_{p=0}^{m\!-\!n\!-\!1}\!\!\!
\left(\!\!\begin{array}{c}
		\!m\!-\!n\!-\!1\!\\
		p
	\end{array}\!\!\right)\!\frac{(-1)^p}{n+p} \Big( \frac{\varpi_4^m}{m} \\ &  +\sum\limits_{q=0}^{n+p}\!\!
\left(\!\!\begin{array}{c}
		\!n\!+\!p\!\\
		q \end{array}\!\!\right)\!\frac{(1\!-\!2\beta)^q(1\!-\!\beta)^{n+p-q} (\epsilon_m^{m-q}\!\!-\!\varpi_4^{m-q})}{(\beta^2\eta)^{n+p}(m-q)} \Big).\notag
\end{align}
\begin{IEEEproof}
	Please refer to Appendix E.	
\end{IEEEproof}
\end{Corollary}

\begin{Remark}
From the results shown in Corollary $3$, it is not hard to find that:
for the case where $m>n$, when $\rho_n \to \infty$, $\rho_m \to \infty$, and $\frac{\rho_n}{\rho_m}=\eta$ is a constant,  $\tilde{P}_n$ can approach zero under the conditions that $\epsilon_m\leq\frac{\beta}{1-\beta}$ or $\frac{\rho_n}{\rho_m}\leq\frac{1}{\beta\epsilon_m}$. Note that, when $m>n$, $\bar{P}_n$ cannot approach zero in FSIC aided hybrid NOMA, which again verifies the benefit of applying HSIC. Another interesting observation is that
the conditions for $\tilde{P}_n$ approaching zero are different for $m>n$ and $m<n$.
\end{Remark}

\begin{Remark}
When $m>n$, $\rho_n \to \infty$, $\rho_m \to \infty$, and $\frac{\rho_n}{\rho_m}$ is a constant,
it is not hard to observe that $\tilde{P}_n$ is proportional to $\frac{1}{\rho_m^n}$ for $\epsilon_m\leq\frac{\beta}{1-\beta}$,
and  $\frac{1}{\rho_n^n}$ for $\frac{\rho_n}{\rho_m}\leq\frac{1}{\beta\epsilon_m}$, which means that
$\tilde{P}_n$ decays with exponential rate $n$ at high SNRs. Thus, the impact of the value of $n$ on $\tilde{P}_n$ is still more dominant compared to that of the value of $m$, which is consistent with conclusions for $m<n$.
\end{Remark}

Similar to Corollaries $2$ and $3$, for $n<m$, the behavior of $\tilde{P}_n$ in another two limiting cases,
i.e., when  $\rho_m\to\infty$ while $\rho_n$ is a constant and when $\rho_n\to\infty$ while $\rho_m$ is a constant,
can be characterized as highlighted in the following two corollaries.

\begin{Corollary}
For the case that $\rho_m\to\infty$ and $\rho_n$ is a constant, the probability $\tilde{P}_n$ for $m>n$ approaches a non-zero constant, which can be expressed as follows:
\begin{align}
\tilde{P}_n \approx & \hat{c}_{mn}\!\frac{\sum\limits_{p=0}^{m\!-\!n\!-\!1}\!\!\hat{c}_p\!\sum\limits_{l=0}^{n\!-\!1}\!\hat{c}_l}{l+p+1}
\!\!\left(\frac{1}{M-n-p}-\frac{1}{M-n+l+1}\right)\notag \\
&\times \left(1-e^{-(M-n+l+1)\frac{1-2\beta}{\beta^2\rho_n}}\right).
\end{align}
\end{Corollary}

\begin{Corollary}
When  $\rho_n\to\infty$ and $\rho_m$ is a constant, $\tilde{P}_n$ for $m>n$ can be approximated as follows:
 \begin{align}
\tilde{P}_n\approx &\frac{1}{\rho_n^n}\hat{c}_{mn}\!\!\!\sum\limits_{s=0}^{m\!-\!n\!-\!1}\!\!\!
\left(\!\!\begin{array}{c}
		\!m\!-\!n\!-\!1\!\\
		s
	\end{array}\!\!\right)\!\frac{(-1)^s}{n} e^{-(M-m+1+s)\alpha_m}\notag\\
&\Big(\sum\limits_{i=1}^{n}\!
\left(\!\!\begin{array}{c}
		n\\
		i
	\end{array}\!\!\right)\!\Big( \frac{(1-\beta)^i\rho_m^i(1-2\beta)^{n-i}}{\beta^{2n}} -\frac{(-1)^{n-i}\alpha_m^{-i}}{\beta^n}  \Big)\notag\\
&\times \Big( \frac{\alpha_m^i}{M\!-\!m\!+\!1\!+\!s}\!+\!\sum\limits_{q=1}^{i}\!\frac{i! \alpha_m^{i-q}}{(i\!-\!q)!(M\!-\!m\!+\!1\!+\!s)^{q+1}}  \Big)\notag\\
&+ \Big((\frac{1-2\beta}{\beta^2})^n-(\frac{-1}{\beta})^n\Big)\frac{1}{M-m+1+s}\Big)\notag\\
&+\frac{1}{\rho_n^n}\hat{c}_n\frac{1}{n}(\frac{1-2\beta}{\beta^2})^n,
\end{align}
when $\epsilon_m>\frac{\beta}{1-\beta}$ and
\begin{align}
\tilde{P}_n\approx \frac{1}{\rho_n^n}\hat{c}_n\frac{1}{n}(\frac{1-2\beta}{\beta^2})^n.
\end{align}
when $\epsilon_m\le\frac{\beta}{1-\beta}$.
\end{Corollary}

It can be seen from the above
corollary that $\tilde{P}_n$ is proportional to $\frac{1}{\rho_n^n}$
when $\rho_n \to \infty$ and $\rho_m$ is a constant,
for both cases with $\epsilon_m>\frac{\beta}{1-\beta}$ and $\epsilon_m\le\frac{\beta}{1-\beta}$.
Hence, it can be concluded that $\tilde{P}_n$ can also avoid probability floors
when $\rho_n \to \infty$ and $\rho_m$ is a constant.

{\color{black}By summarizing the results shown in the above corollaries and remarks, the following conclusions
can be obtained which may be useful for practical designs:
\begin{itemize}
	\item when $\rho_m$ is a constant, it is favorable to have a large $\rho_n$, for both cases $m>n$ and $m<n$;
	\item when $\rho_n$ is a constant, there is a floor of $\tilde{P}_n$, even with a sufficiently large $\rho_m$,
	      for both cases $m>n$ and $m<n$;
    \item when both the transmit powers of $U_m$ and $U_n$ becomes large, it is always preferable to have a small
    target data rate of $U_m$ so that $\epsilon_m\leq\frac{\beta}{1-\beta}$, for both cases $m>n$ and $m<n$.
	However, when $R_m$ becomes large so that  $\epsilon_m>\frac{\beta}{1-\beta}$, whether $\tilde{P}_n$ can approach
	zero depends on the relationship between the transmit powers of the two users. Specifically,
	to make $\tilde{P}_n$ approach zero, $\frac{\rho_n}{\rho_m}$ should be larger than $\frac{1-\beta}{\beta^2}$ when
	$m<n$, while $\frac{\rho_n}{\rho_m}$ should be less than or equal to $\frac{1}{\epsilon_m\beta}$ when
	$m>n$. The above observation indicates that the relationship of the channel gains of the
	two users play a subtle role on the performance.
\end{itemize}
}

\section{Numerical Results}
\begin{figure}[!t]
	\centering
	\vspace{0em}
	\setlength{\abovecaptionskip}{0em}   
	\setlength{\belowcaptionskip}{-2em}   
\subfloat[$m<n$, $\frac{\rho_n}{\rho_m}=5$, $m=1$ ]{\includegraphics[width=3.2in]{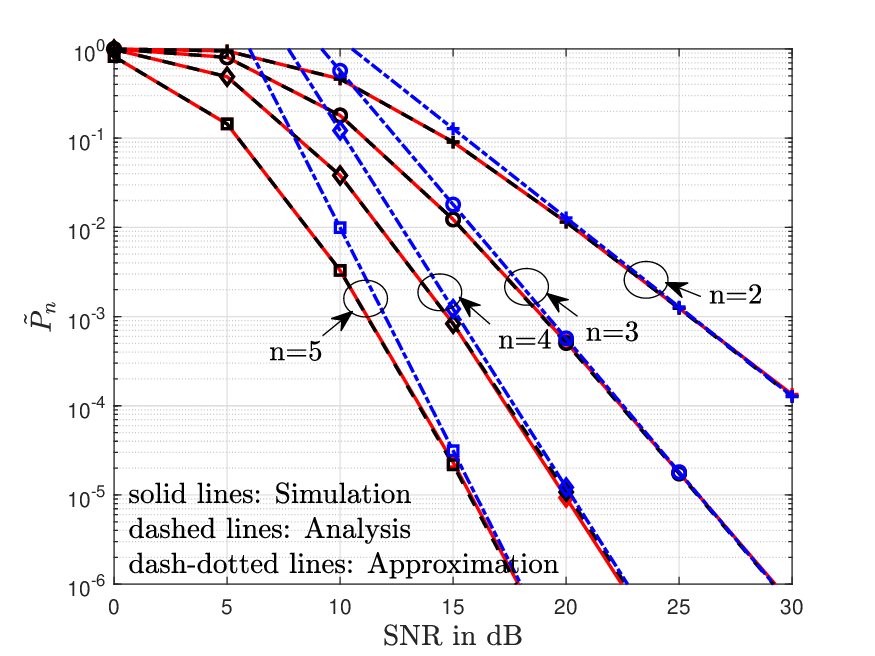}%
\label{fig1_a}}
\\
\subfloat[$m>n$, $\frac{\rho_n}{\rho_m}=5$, $m=5$]{\includegraphics[width=3.2in]{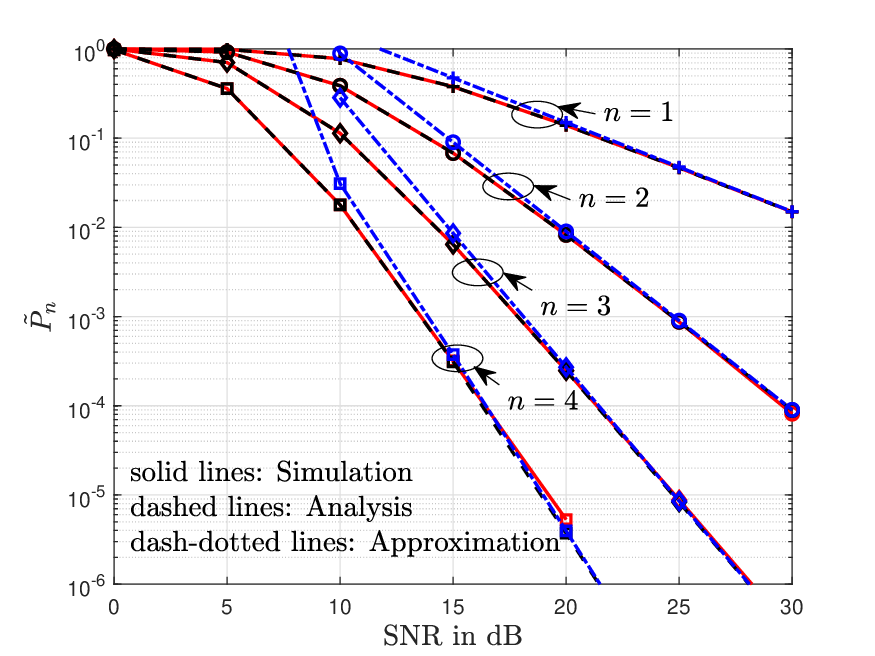}%
\label{fig1_b}}
	\caption{$\tilde{P}_n$ versus SNR for $m>n$ and $m<n$. $R_m=0.2$ bits per channel use (BPCU), $M=5$, and $\beta=\frac{1}{3}$.}
	\label{fig1}
\end{figure}
In this section, simulation results are provided to verify the accuracy of the developed analysis and demonstrate the performance of the proposed HSIC scheme.
Comparisons with the benchmark FSIC scheme are also provided.
Besides, it will be shown that user pairing has a significant impact on the performance of the proposed HSIC scheme.

\begin{figure}[!t]
	\centering
	\vspace{0em}
	\setlength{\abovecaptionskip}{0em}   
	\setlength{\belowcaptionskip}{-2em}   
\subfloat[$m<n$, $\frac{\rho_n}{\rho_m}=7$, $n=5$, $R_m=1$ ]{\includegraphics[width=3.2in]{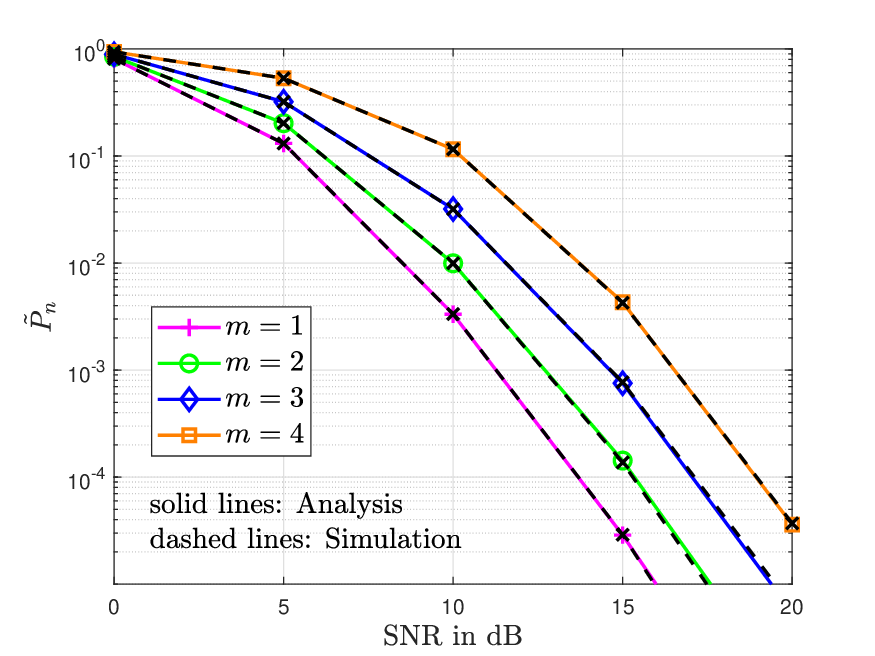}%
\label{fig2_a}}
\\
\subfloat[$m>n$, $\frac{\rho_n}{\rho_m}=7$,  $n=1$, $R_m=0.2$ ]{\includegraphics[width=3.2in]{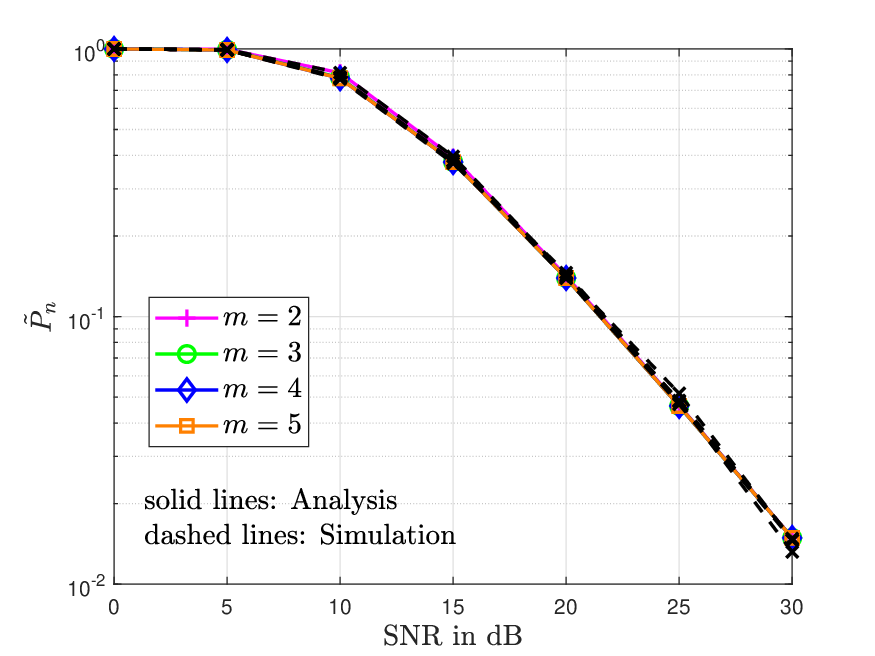}%
\label{fig2_b}}
	\caption{$\tilde{P}_n$ with varying $m$ and fixed $n$. $M=5$ and $\beta=\frac{1}{3}$.}
	\label{fig2}
\end{figure}

\begin{figure}[!t]
	\centering
	\vspace{0em}
	\setlength{\abovecaptionskip}{0em}   
	\setlength{\belowcaptionskip}{0em}   
\subfloat[$m<n$, $m=2$, $n=5$, $R_m=0.2$, $\epsilon_m<\frac{\beta}{1-\beta}$ ]{\includegraphics[width=3.2in]{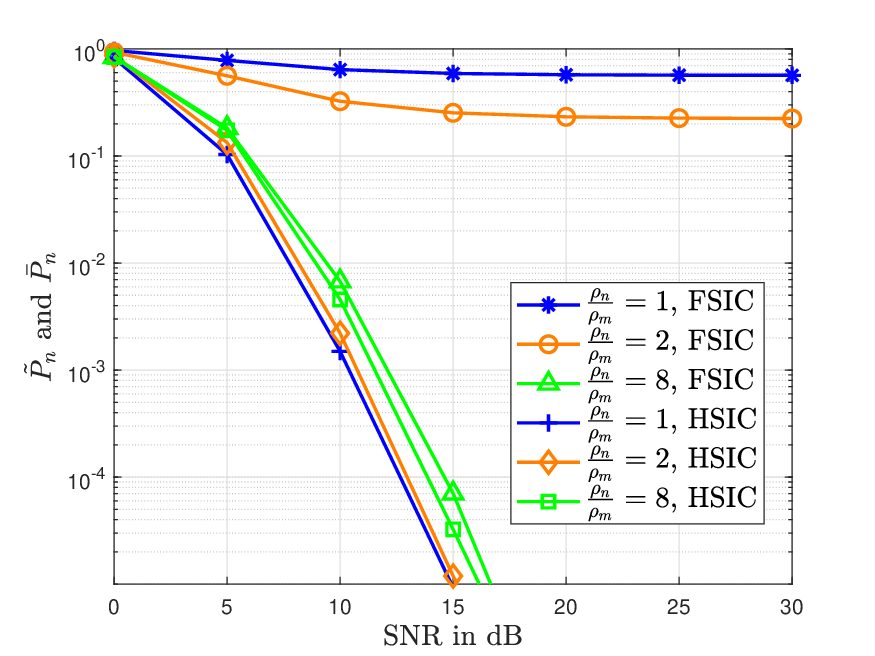}%
\label{fig3_a}}
\\
\subfloat[$m>n$, $m=5$, $n=2$, $R_m=1$, $\epsilon_m>\frac{\beta}{1-\beta}$]{\includegraphics[width=3.2in]{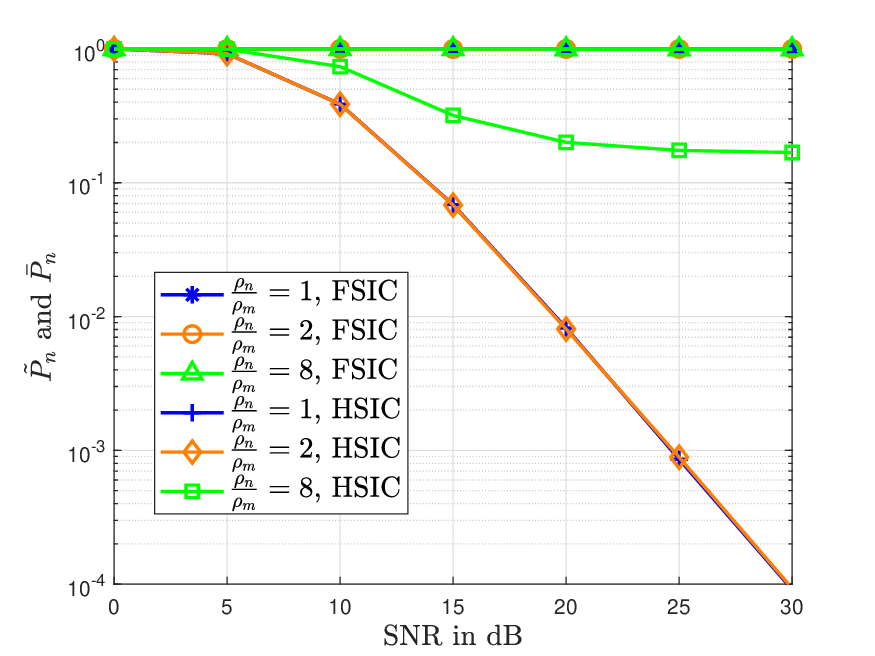}%
\label{fig3_b}}
	\caption{Comparisons of HSIC and FSIC aided hybrid NOMA schemes in terms of $\tilde{P}_n$ and $\bar{P}_n$. $M=5$, $\beta=\frac{1}{3}$.}
	\label{fig3}
\end{figure}

Fig. \ref{fig1} shows $\tilde{P}_n$ versus SNR for $m>n$ and $m<n$ in Fig. \ref{fig1} (a) and Fig. \ref{fig1} (b), respectively \footnote{Note that in this section, $\text{SNR}=\rho_n$, since the noise power is normalized.}. Note that the curves for analytical results are based on Theorems $1$ and $2$, respectively, while those for approximations are based on Corollaries $1$ and $4$, respectively.
As shown in Figs. \ref{fig1} (a) and (b), analytical results perfectly match simulations,
as well as the approximations at high SNRs, which verifies the accuracy of the developed analysis.
As shown in the figure, when $m$ is fixed, the value of $n$ has a significant impact on $\tilde{P}_n$.
For example,  Figs. \ref{fig1} (a), in for the case where $m<n$ and $m=1$, $\tilde{P}_n$ for $n=5$ is the lowest among all choices of $n$. Besides, it can be observed that the slope of the curve for $n=5$ is the largest, which is consistent with the conclusion stated in Remark 2.

Fig. \ref{fig2} shows the impact of the value of $m$ on $\tilde{P}_n$ for a fixed $n$.
From Fig. \ref{fig2} (a), it can be straightforwardly observed that when $m<n$ $\tilde{P}_n$ decrease with $m$.
However, at high SNRs, the decaying rates of $\tilde{P}_n$ (i.e., the slopes of the curves )
for different values of $m$ are almost same. This observation is consistent
with Corollary $2$ that the decaying rate of $\tilde{P}_n$ at high SNR is $n$,
instead of $m$. In contrast, it can be seen from Fig. \ref{fig2} (b) that,
when $m>n$, the value of $m$ has little impact on $\tilde{P}_n$.

\begin{figure}[!t]
	\centering
	\vspace{0em}
	\setlength{\abovecaptionskip}{0em}   
	\setlength{\belowcaptionskip}{-1em}   
\subfloat[$m<n$, $m=2$, $n=5$, $R_m=0.2$]{\includegraphics[width=3.2in]{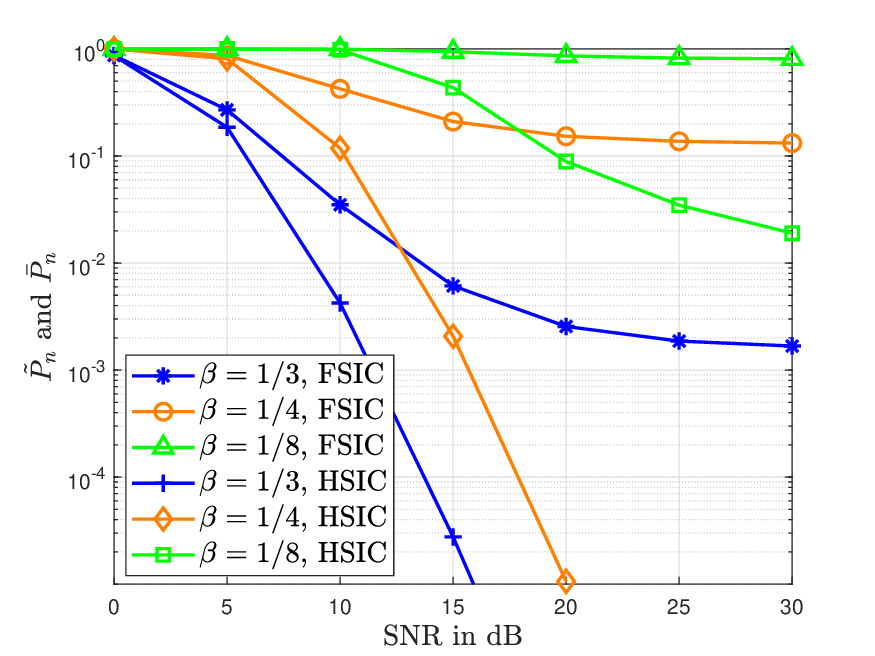}%
\label{fig4_a}}
\\
\subfloat[$m>n$, $m=5$, $n=2$, $R_m=0.2$]{\includegraphics[width=3.2in]{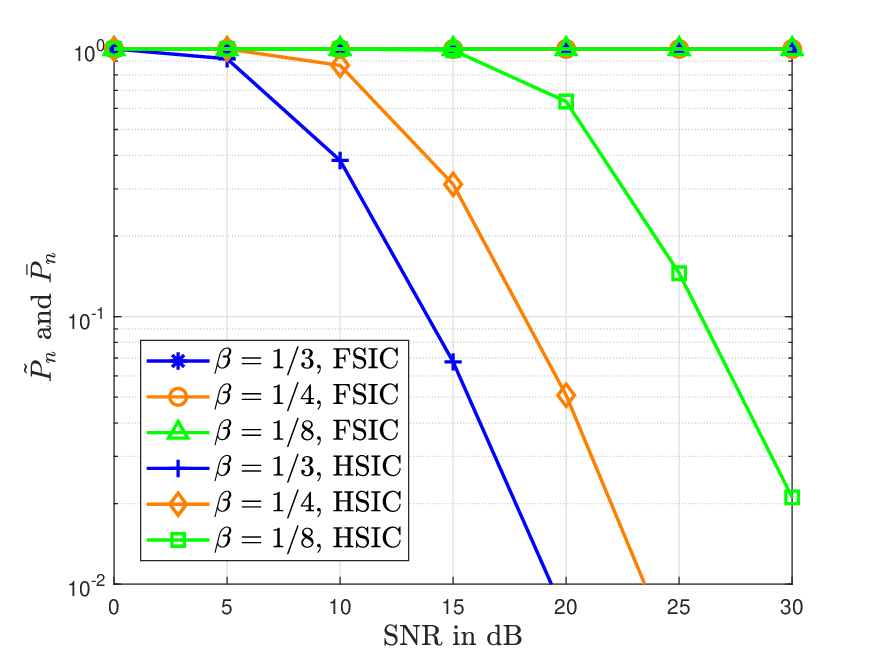}%
\label{fig4_b}}
	\caption{Comparisons of HSIC and FSIC aided hybrid NOMA schemes in terms of $\tilde{P}_n$ and $\bar{P}_n$. $\frac{\rho_n}{\rho_m}=5$, and $M=5$.}
	\label{fig4}
\end{figure}
Fig. \ref{fig3} and Fig. \ref{fig4} compare $\tilde{P}_n$ achieved by the proposed HSIC aided hybrid NOMA with
$\bar{P}_n$ for the existing FSIC aided hybrid NOMA scheme, with varying $\frac{\rho_n}{\rho_m}$ and $\beta$, respectively.
From Fig. \ref{fig3} (a), it can be seen that $\bar{P}_n$ for FSIC can only approach zero when $\frac{\rho_n}{\rho_m}=8$, while $\tilde{P}_n$ for HSIC can approach zero for all values of $\frac{\rho_n}{\rho_m}$. This is because, as stated in Remark 1, $P_n
$ for FSIC aided hybrid NOMA can  approach zero only when $\frac{\rho_n}{\rho_m}>\frac{1-\beta}{\beta^2}$, while $\tilde{P}_n$ can approach zero for all values of $\frac{\rho_n}{\rho_m}$ when $\epsilon_m>\frac{1-\beta}{\beta^2}$. In contrast, when $m>n$, it can be seen from
Fig.  \ref{fig3} (b) that there are severe floors at high SNRs for $\bar{P}_n$ for all choices of $\frac{\rho_n}{\rho_m}$,
while $\tilde{P}_n$ can avoid the floors when $\frac{\rho_n}{\rho_m}=1$ or $2$, which is consistent with the statement 
in Remark $3$. 
Another interesting observation from Fig. \ref{fig3} is that, $\bar{P}_n$ decreases with
$\frac{\rho_n}{\rho_m}$, while $\tilde{P}_n$ increases with $\frac{\rho_n}{\rho_m}$. 
{\color{black}The aforementioned observation indicates that when $\rho_n$ is fixed, increasing the value of 
${\rho_m}$ is favorable for improving the performance of HSIC aided hybrid NOMA. In contrast, 
the increasing of $\rho_m$ will degrade the performance of FSIC aided hybrid NOMA.}
Similar to the observations in Fig. \ref{fig3}, it is shown in Fig. \ref{fig4} that $\bar{P}_n$ can avoid
floors for very limited cases, while there are more cases where $\tilde{P}_n$ can avoid floors. Last but not  least, as shown in both Figs. \ref{fig3} and \ref{fig4}, $\tilde{P}_n$ outperforms $\bar{P}_n$ for most of the cases, which indicates the benefit of applying HSIC to hybrid NOMA.

\begin{figure}[!t]
	\centering
	\vspace{0em}
	\setlength{\abovecaptionskip}{0em}   
	\setlength{\belowcaptionskip}{-2em}   
	\includegraphics[width=3.2in]{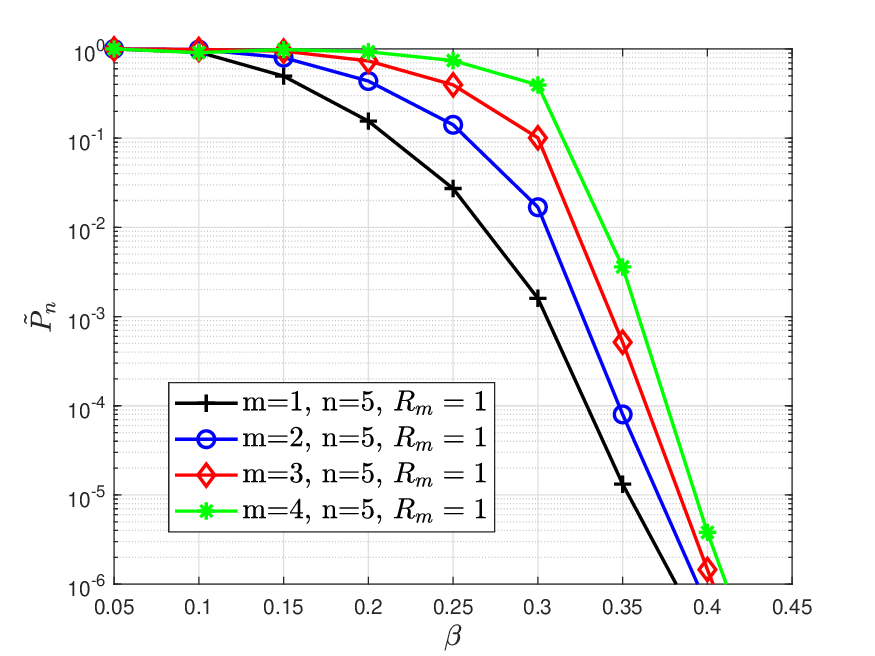}\\
	\caption{$\tilde{P}_n$ versus $\beta$. $R_m=1$, $n=5$, $\frac{\rho_n}{\rho_m}=6$, SNR=$15$ dB.}
	\label{fig5}
\end{figure}

Fig. \ref{fig5} illustrates $\tilde{P}_n$ versus $\beta$.
It can be easily seen from the figure
{\color{black}that} $\tilde{P}_n$  decreases as $\beta$ increases.
This can be straightforwardly explained that a larger $\beta$ can directly yield a larger instantaneous achievable rate for the considered HSIC aided hybrid NOMA scheme. Besides, it is shown that the declining rates
of the curves become faster as $\beta$ increases.
This observation indicates that, after $\beta$ exceeds a threshold,
a slight increase {\color{black}of} the transmit power can result in a significant performance improvement for the proposed scheme.

 \begin{figure}[!t]
	\centering
	\vspace{0em}
	\setlength{\abovecaptionskip}{0em}   
	\setlength{\belowcaptionskip}{-1em}   
	\includegraphics[width=3.2in]{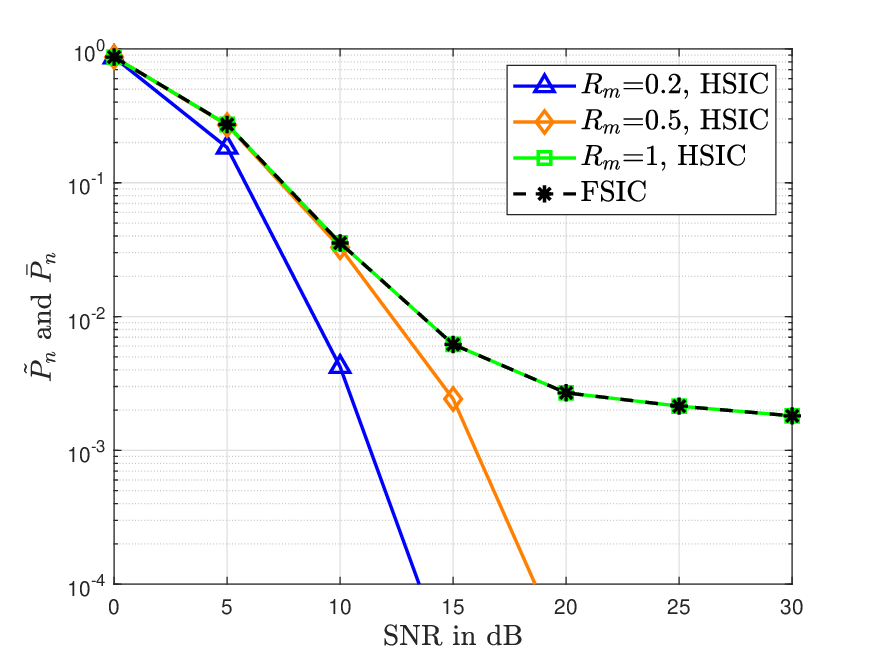}\\
	\caption{Impact of $R_m$ on the comparison between HSIC aided hybrid NOMA and FSIC aided hybrid NOMA. $n=5$, $m=2$, $\frac{\rho_n}{\rho_m}=5$.}
	\label{fig_7_10_29}
\end{figure}
{\color{black}Fig. \ref{fig_7_10_29} shows the impact of the target data rate $R_m$ of the legacy user on
$\tilde{P}_n$ and $\bar{P}_n$, respectively. Note that the variation of $R_m$ has no impact on
$\bar{P}_n$ achieved by FSIC aided hybrid NOMA. This is because in the NOMA transmission phase of
FISC aided hybrid NOMA, the secondary user's signal can be decoded only at the first stage of SIC, so that
the achievable data rate of the opportunistic user is affected by the transmit power and channel gain of the legacy user, instead of $R_m$. Besides, it can be observed from the figure that $\tilde{P}_n$ increases with
$R_m$. The reasons are mainly two fold as follows. First, if the opportunistic user can be decoded at the
second stage of SIC, it can achieve a higher data rate than the case when it is decoded at the first stage of SIC. Second, a larger $R_m$ yields a larger $\tau_m$, which is benefit for the opportunistic user to
be decoded in the second stage of SIC. Furthermore, as $R_m$ becomes sufficiently large, the performance
of HSIC aided hybrid NOMA will be the same as that of the FSIC counterpart. 
Thus, in practice, the proposed HSIC aided hybrid NOMA scheme favors to be applied in the
scenarios where the legacy users have low target data rate.}

\section{Conclusion}
In this paper, a novel HSIC aided hybrid NOMA scheme has been proposed, which can be implemented as a simple add-on to the legacy OMA network. A power coefficient has been introduced to ensure that the energy consumption of the proposed scheme is less than that of OMA. Closed-form expression for the probability $\tilde{P}_n$ that the achievable rate of the proposed HSIC aided hybrid NOMA scheme cannot outperform its OMA counterpart has been obtained. It has been shown by asymptotic results as well as numerical results that HSIC aided hybrid NOMA outperforms its FSIC counterpart.
Specifically, it has been shown that the conditions for $\tilde{P}_n$ approaches zero by applying HSIC can be more relaxed compared to FSIC. The impact of user pairing has also been considered, which indicates that the order of the opportunistic user plays more significant role on the performance than that of its paired legacy user.

{\color{black}In this paper, slow time varying channel has been considered and
it has been assumed that the channel gain of a user in a frame remains a constant.
In future, it is very important to investigate the application of
HSIC in time varying hybrid NOMA-TDMA and hybrid NOMA-FDMA scenarios. Besides, perfect CSI
has been assumed in the paper, it is also important to study the performance of
HSIC aided hybrid NOMA under imperfect CSI conditions.}

\appendices
\section{Proof for Lemma 1}
The evaluation of $\tilde{P}_n$ can be divided into two cases, one is $\beta\rho_n|h_n|^2\le\tau(m)$, and the other is $\beta\rho_n|h_n|^2>\tau(m)$. As a result, $\tilde{P}_n$ can be rewritten as follows:
\begin{align}
\tilde{P}_n=&\underbrace{P\left( T\log(1+\beta\rho_{ n}\left | h_{ n}  \right |^2)+T\log(1+\beta \rho_n |h_n|^2)\le\right.}_{P_1} \notag\\
& \underbrace{\left.T\log(1+\rho_n|h_n|^2) ,  \beta \rho_{n}\left | h_{n} \right |^2 \le  \tau_m\right)}_{P_1}        \notag\\
&+\underbrace{P\left(\log(1\!+\!\frac{\beta \rho_{n}\left | h_{n} \right |^2 }{\rho_{m}\left | h_m \right | ^2\!+\!1} )+T\log(1\!+\!\beta \rho_n |h_n|^2)\le\right.}_{P_2} \notag\\
&\underbrace{\left.T\log(1+ \rho_n |h_n|^2) , \beta \rho_{n}\left | h_{n} \right |^2 > \tau_m\right)}_{P_2}.
\end{align}

For $P_1$, with some algebraic manipulations, it can be expressed as follows:
\begin{align}
P_1=&P\left( (1+\beta\rho_n|h_n|^2)^2\le 1+\rho_n|h_n|^2,|h_m|^2>\alpha_m,\right.\notag\\
&\left. |h_n|^2 \le \Phi(|h_m|^2)
  \right)\\
=&P\!\left(\! |h_n|^2\!\le \!\frac{1\!-\!2\beta}{\beta^2\rho_n} ,|h_m|^2\!>\!\alpha_m, |h_n|^2 \!\le \! \Phi(|h_m|^2)\right),\notag
\end{align}
where the first step is obtained by the fact that $\tau(m)=0$ when $|h_m|^2<\alpha_m$, which makes $P_1=0$.
By noting that if $|h_m|^2< \frac{1-\beta}{\beta}\alpha_m$, $\frac{1-2\beta}{\beta^2\rho_n}>\Phi(|h_m|^2)$ holds, otherwise $\frac{1-2\beta}{\beta^2\rho_n}<\Phi(|h_m|^2)$ holds. Thus, $P_1$ can be rewritten as follows:
\begin{align}
P_1=&\underbrace{P\left( |h_n|^2\le \Phi(|h_m|^2),\alpha_m<|h_m|^2< \omega_2\right)}_{P_{1,1}}\notag\\
&+\underbrace{P\left(  |h_n|^2\le \omega_3, |h_m|^2> \omega_2 \right)}_{P_{1,2}}.
\end{align}

According to whether $\tau(m)=0$, $P_2$ can be divided into the sum of two parts as follows:
\begin{align}
P_2=&\underbrace{P\left( |h_n|^2\le \Psi(|h_m|^2), |h_n|^2>\Phi(|h_m|^2)
,|h_m|^2>\alpha_m \right)}_{P_{2,1}}\notag\\
&+\underbrace{P\left(  |h_n|^2\le \Psi(|h_m|^2), |h_m|^2<\alpha_m  \right)}_{P_{2,2}},
\end{align}

Therefore, $\tilde{P}_n=P_{1,1}+P_{1,2}+P_{2,1}+P_{2,2}$ and the proof is complete.

\section{Proof for Theorem 1 ($m<n$)}
{\color{black}
According to Lemma $1$, $\tilde{P}_n=P_{1,1}+P_{1,2}+P_{2,1}+P_{2,2}$.
The next section will only provide the proof for $P_{1,1}$, since the proofs for the other three parts are similar to the proof for $P_{1,1}$.

For $P_{1,1}$, note that there is a hidden condition that $|h_m|^2<\Phi(|h_m|^2)$ since $|h_m|^2<|h_n|^2$.
If $\frac{\rho_n}{\rho_m}>\frac{\epsilon_m^{-1}}{\beta}$, $\Phi(|h_m|^2)<|h_m|^2$ always holds.
If $\frac{\rho_n}{\rho_m}<\frac{\epsilon_m^{-1}}{\beta}$, when $|h_m|^2>\frac{1}{\alpha_m^{-1}-\beta\rho_n}$, $\Phi(|h_m|^2)
>|h_m|^2$ holds, otherwise $\Phi(|h_m|^2)
<|h_m|^2$.

Therefore, when $\frac{\rho_n}{\rho_m}>\frac{\epsilon_m^{-1}}{\beta}$, $P_{1,1}=0$, and when $\frac{\rho_n}{\rho_m}<\frac{\epsilon_m^{-1}}{\beta}$, $P_{1,1}$ can be rewritten as follows:
\begin{align}
P_{1,1}
=&P\left(|h_m|^2\!\le\! |h_n|^2\!\le\! \Phi(|h_m|^2),\omega_1\!<\!|h_m|^2\!<\! \omega_2\right).
\end{align}
Note that when $\frac{\rho_n}{\rho_m}<\kappa_1$, the hidden condition that $\omega_1<\omega_2$ always holds. Thus, when $\kappa_1<\frac{\rho_n}{\rho_m}<\frac{\epsilon_m^{-1}}{\beta}$, $P_{1,1}=0$.
}

 Note that the users are ordered according to their channel gains, and hence
 the joint probability density function (pdf) of $|h_m|^2$ and $|h_n|^2$ $(m<n)$ is given by:
 \begin{align}
&f_{|h_{m}|^2,|h_n|^2}(x,y)\notag\\
=&c_{mn}\!\!\!\sum_{p=0}^{n-m-1}\!\!\!\!c_p\!\!\sum_{l=0}^{m-1}\!c_l e^{-(l+p+1)x}e^{-(M-m-p)y},
 \end{align}

 Thus, when $\frac{\rho_n
 }{\rho_m}<\kappa_c$, $P_{1,1}$ can be expressed as follows :
\begin{align}\label{P_{1,1}}
P_{1,1}=&\int_{\omega_1}^{\omega_2}\int_{x}^{\Phi(x)}
 c_{mn}\!\!\!\!\sum_{p=0}^{n-m-1}\!\!\!\!c_p\!\!\sum_{l=0}^{m-1}\!c_l e^{-(l+p+1)x}\times\notag\\
 &e^{-(M-m-p)y}dydx\notag\\
 =&c_{mn}\!\!\!\sum_{p=0}^{n-m-\!1}\!\!\!c_p\!\!\sum_{l=0}^{m-1}\!c_l\frac{1}{M-m-p} \times \notag\\ &\Big(\frac{e^{-(M-m+l+1)\omega_1}-e^{-(M-m+l+1)\omega_2}}{M-m+l+1}   - \notag\\
 & e^{\frac{M-m-p}{\beta\rho_n}}\frac{e^{-r\omega_1}-e^{-r\omega_2}}{r}\Big),
\end{align}
where $\Phi(x)=\frac{\alpha_m^{-1}x-1}{\beta\rho_n}$.

Therefore, $P_{1,1}$ can be expressed as (\ref{P_{1,1}}) when $\frac{\rho_n}{\rho_m}<\kappa_1$, and $P_{1,1}=0$ when $\frac{\rho_n}{\rho_m}>\kappa_1$.

Finally, similar to the proof for $P_{1,1}$ above, the expression for $P_{1,2}$, $P_{2,1}$ and $P_{2,2}$ can be obtained and the proof is complete.
\section{Proof for Corollary $1$ ($m<n$)}
Consider that the proof steps for $P_1$, $P_{2,1}$, and $P_{2,2}$ are similar,
only the proof for $P_1$ will be provided in the following.

Recall that $P_1=P_{1,1}+P_{1,2}$. In the following, the approximation for $P_{1,1}$ is obtained first, and then the approximation for $P_{1,2}$ is developed.

First, recall that when $\frac{\rho_n}{\rho_m} \le \kappa_1$, $P_{1,1}$ can be expressed as:
\begin{align}
P_{1,1}
=&P\left(|h_m|^2\!\le\! |h_n|^2\!\le\! \Phi(|h_m|^2),\omega_1\!<\!|h_m|^2\!<\! \omega_2\right).
\end{align}
In order to facilitate the approximation, the pdf of $|h_m|^2$ and $|h_n|^2$ ($m<n$) can be rewritten as follows:
 \begin{align}
&f_{|h_{m}|^2,|h_n|^2}(x,y)\\
=&c_{mn}\!\left( F(x) \right)^{m\!-\!1}\!\!\left(  F(y) \!-\!F(x) \right)^{n\!-\!m\!-\!1}\!\!\left( 1\!-\!F(y) \right)^{M\!-\!n}\!\!f(x)f(y),\notag
 \end{align}
where $x<y$, $f(x)=e^{-x}$, $F(x)=1-e^{-x}$.
By applying the above pdf, $P_{1,1}$ can be rewritten as follows:
\begin{align}
P_{1,1}
=&\!\!\underbrace{\int_{\omega_1}^{\omega_3}\!\!\!\!\int_{(\beta\rho_ny+1)\alpha_m}^{y}f_{|h_{m}|^2,|h_n|^2}(x,y)dx dy}_{\Gamma_1}\notag\\
&-\!\!\underbrace{\int_{\omega_2}^{\omega_3}\!\!\!\!\int_{\omega_2}^{y}f_{|h_{m}|^2,|h_n|^2}(x,y)dx dy}_{\Gamma_2}.
\end{align}
 By applying the binomial expansion, the joint pdf can be further rewritten as:
 \begin{align}
&f_{|h_{m}|^2,|h_n|^2}(x,y)=c_{mn}\!\sum\limits_{p=0}^{n\!-\!m\!-\!1}\!\!\!
\left(\!\!\begin{array}{c}
		n\!-\!m\!-\!1\\
		p
	\end{array}\!\!\right)\!(-1)^p f(y) \\
&\times \left(F(y)\right)^{n-m-1-p}\left(1-F(y)\right)^{M-n}f(x)\left(F(x)\right)^{m-1+p}   ,\notag
 \end{align}
 {\color{black} Therefore, the first term of $P_{1,1}$, $\Gamma_1$, can be evaluated as follows:}
 \begin{align}
 &\Gamma_1=c_{mn}\!\sum\limits_{p=0}^{n\!-\!m\!-\!1}\!\!\!
\left(\!\!\begin{array}{c}
		n\!-\!m\!-\!1\\
		p
	\end{array}\!\!\right)\!(-1)^p \int_{\omega_1}^{\omega_3}\!\!\!\!\int_{(\beta\rho_ny+1)\alpha_m}^{y}\!\!\!\!f(y)\\
&\times \left(F(y)\right)^{n-m-1-p}\!\left(1-F(y)\right)^{M-n}\!\!f(x)\left(F(x)\right)^{m-1+p}dx dy \notag\\
&=c_{mn}\!\sum\limits_{p=0}^{n\!-\!m\!-\!1}\!\!\!
\left(\!\!\begin{array}{c}
		n\!-\!m\!-\!1\\
		p
	\end{array}\!\!\right)\!\frac{(-1)^p }{m+p} \int_{\omega_1}^{\omega_3}\!\!\!\!f(y)\left(F(y)\right)^{n-m-1-p}\!\!\times\notag\\
&\left(1\!-\!F(y)\right)^{\!M\!-\!n}\!\!\Big((F(y))^{m+p}\!\!-\!(F((\beta\rho_ny\!+\!1)\alpha_m))^{m+p}\Big)dy\notag.
\end{align}
By applying Taylor seizes, the two functions $f(x)$ and $F(x)$ can be approximated as follows: $f(x)=e^{-x}\approx 1$ and $F(x)=1-e^{-x} \approx x$ when $x\to 0$.
Therefore, $\Gamma_1$ can be further approximated as follows:
\begin{align} \label{gamma_1}
\Gamma_1\approx& c_{mn}\!\sum\limits_{p=0}^{n\!-\!m\!-\!1}\!\!\!
\left(\!\!\begin{array}{c}
		n\!-\!m\!-\!1\\
		p
	\end{array}\!\!\right)\!\frac{(-1)^p }{m+p} \int_{\omega_1}^{\omega_3}\!\!y^{n-m-1-p}\notag\\
&\times \Big( y^{m+p}\!\!-\!((\beta\rho_ny\!+\!1)\alpha_m)^{m+p}\Big)dy\notag\\
\overset{(a)}{=}& c_{mn}\!\sum\limits_{p=0}^{n\!-\!m\!-\!1}\!\!\!
\left(\!\!\begin{array}{c}
		n\!-\!m\!-\!1\\
		p
	\end{array}\!\!\right)\!\frac{(-1)^p }{m+p} \int_{\omega_1}^{\omega_3}\Big(y^{n-1} -\!\!\notag \\
&  \alpha_m^{m+p} \sum_{q=0}^{m+p}\!\!\!
\left(\!\!\begin{array}{c}
		m+p\\
		q
	\end{array}\!\!\right)\left( \beta \rho_n \right)^{m+p-q} y^{n-1-q}   \Big)dy \notag\\
=& \frac{1}{\rho_m^n}c_{mn}\!\sum\limits_{p=0}^{n\!-\!m\!-\!1}\!\!\!
\left(\!\!\begin{array}{c}
		n\!-\!m\!-\!1\\
		p
	\end{array}\!\!\right)\!\frac{(-1)^p }{m+p} \Big(\frac{\varpi_3^n-\varpi_1^n}{n} -\!\! \\
&  \epsilon_m^{m+p} \sum_{q=0}^{m+p}\!\!\!
\left(\!\!\begin{array}{c}
		m+p\\
		q
	\end{array}\!\!\right)\left( \beta \eta \right)^{m+p-q} \frac{\varpi_3^{n-q}-\varpi_1^{n-q}}{n-q}   \Big),\notag
\end{align}
where the step (a) is obtained by applying the binomial expansion.

The steps for the derivation of $\Gamma_2$ are similar to those of $\Gamma_1$, yielding the following approximation:
\begin{align} \label{gamma_2}
\Gamma_2\approx&  \frac{1}{\rho_m^n}c_{mn}\!\sum\limits_{p=0}^{n\!-\!m\!-\!1}\!\!\!
\left(\!\!\begin{array}{c}
		n\!-\!m\!-\!1\\
		p
	\end{array}\!\!\right)\!\frac{(-1)^p }{m+p} \Big(\frac{\varpi_3^n-\varpi_2^n}{n} -\!\! \notag\\
&  \varpi_2^{m+p}  \frac{\varpi_3^{n-m-p}-\varpi_2^{n-m-p}}{n-m-p} \Big).
\end{align}

Similar to the above steps, when $\frac{\rho_n}{\rho_m} \le \kappa_1$, $P_{1,2}$ can be approximated as:
\begin{align} \label{P_12}
P_{1,2}\approx&  \frac{1}{\rho_m^n}c_{mn}\!\sum\limits_{p=0}^{n\!-\!m\!-\!1}\!\!\!
\left(\!\!\begin{array}{c}
		n\!-\!m\!-\!1\\
		p
	\end{array}\!\!\right)\!\frac{(-1)^p }{m+p} \Big(\frac{\varpi_3^n-\varpi_2^n}{n} -\!\! \notag\\
&  \varpi_2^{m+p}  \frac{\varpi_3^{n-m-p}-\varpi_2^{n-m-p}}{n-m-p}   \Big).
\end{align}

Then, by combining (\ref{gamma_1}), (\ref{gamma_2}) and (\ref{P_12}), when $\frac{\rho_n}{\rho_m} \le \kappa_1$, $P_1$ can be approximated as follows:
\begin{align}
&P_1  \approx \frac{1}{\rho_m^n}\tilde{T}_1 = \frac{1}{\rho_m^n}c_{mn}\!\!\sum\limits_{p=0}^{n\!-\!m\!-\!1}\!\!\!
\left(\!\!\begin{array}{c}
		\!n\!-\!m\!-\!1\!\\
		p
	\end{array}\!\!\right)\!\frac{(-1)^p }{m+p} \Big(\frac{\varpi_3^n-\varpi_1^n}{n}\!\! \notag\\
& -\epsilon_m^{m+p} \sum_{q=0}^{m+p}\!\!\!
\left(\!\!\begin{array}{c}
		m+p\\
		q
	\end{array}\!\!\right)\left( \beta \eta \right)^{m+p-q} \frac{\varpi_3^{n-q}-\varpi_1^{n-q}}{n-q}   \Big).
\end{align} 
By following the same method for the approximation for $P_1$, the approximations for $P_{2,1}$ and $P_{2,2}$ can be obtained. Hence, the proof is complete.

\section{Proof for Theorem 2 ($m>n$)}
{\color{black}
According to Lemma $1$, the evaluation of $\tilde{P}_n$ can be divided into four parts: $P_{1,1}$,
$P_{1,2}$, $P_{2,1}$ and $P_{2,2}$.
Since the proofs for the four parts are similar,
the next section will only focus on the proof for $P_{1,1}$, and the proofs for the other three parts are similar to the proof for $P_{1,1}$.}

Note that $P_{1,1}$ can be expressed as follows:
\begin{align}
P_{1,1}\!=\!P\left( |h_n|^2\!\le\! \Phi(|h_m|^2),\alpha_m\!<\!|h_m|^2\!<\! \omega_2\right).
\end{align}
Note that if $\frac{\rho_n}{\rho_m}>\frac{\epsilon_m^{-1}}{\beta}$, $\Phi(|h_m|^2)<|h_m|^2$ always holds.
If $\frac{\rho_n}{\rho_m}<\frac{\epsilon_m^{-1}}{\beta}$, when $|h_m|^2<\omega_1$, $\Phi(|h_m|^2)
<|h_m|^2$ holds, otherwise $\Phi(|h_m|^2)>|h_m|^2$.
In addition,
if $\frac{\rho_n}{\rho_m}>\kappa_1$, $\omega_1>\omega_2$,
otherwise $\omega_1<\omega_2$.

When $\frac{\rho_n}{\rho_m}<\kappa_1$, $P_{1,1}$ can be rewritten as follows:
\begin{align}
P_{1,1}
=&P\!\left(|h_n|^2\!\le \!
\Phi(|h_m|^2),\alpha_m\!<\!|h_m|^2\!<\!\omega_1\right)\\
&+P\!\left(  |h_n|^2\!\le \! |h_m|^2,\omega_1\!<\!|h_m|^2\!<\!\omega_2\right)\!.\notag
\end{align}

Since the users are ordered according to their channel gains,
the joint pdf of $|h_n|^2$ and $|h_m|^2$ $(n < m)$ is given by:
 \begin{align}
&f_{|h_{n}|^2,|h_m|^2}(x,y)\notag\\
=&\hat{c}_{mn}\!\!\!\sum_{p=0}^{m-n-1}\!\!\!\!\hat{c}_p\!\!\sum_{l=0}^{n-1}\!\hat{c}_l e^{-(l+p+1)x}e^{-(M-n-p)y}.
 \end{align}

Thus, when $\frac{\rho_n}{\rho_m}<\kappa_1$, $P_{1,1}$ can be expressed as follows :
\begin{align}\label{sum11}
P_{1,1}=&\int_{\alpha_m}^{\omega_1}\int_{0}^{\Phi(y)}
 f_{|h_{n}|^2,|h_m|^2}(x,y)dx dy\notag\\
 &+\int_{\omega_1}^{\omega_2}\int_{0}^{y}
 f_{|h_{n}|^2,|h_m|^2}(x,y)dx dy\notag\\
 =&\hat{c}_{mn}\!\!\!\sum_{p=0}^{m-n-\!1}\!\!\!\hat{c}_p\!\!\sum_{l=0}^{n-1}\!\hat{c}_l\frac{1}{l+p+1} \times \notag\\ &\Big(\frac{e^{-(M-n-p)\alpha_m}-e^{-(M-n-p)\omega_2}}{M-n-p}   \notag\\
 & -e^{\frac{l+p+1}{\beta\rho_n}}\frac{e^{-t\alpha_m}-e^{-t\omega_1}}{t} \notag\\
&  - \frac{e^{-(M-n+l+1)\omega_1}-e^{-(M-n+l+1)\omega_2 }}{M-n+l+1}   \Big).
\end{align}

When $\frac{\rho_n}{\rho_m}>\kappa_1$, $P_{1,1}$ can be rewritten as follows:
\begin{align}\label{sum12}
P_{1,1}
=&P\!\left(|h_n|^2\!\le \! \Phi(|h_m|^2)
 ,\alpha_m\!<\!|h_m|^2\!<\!\omega_2\right)\notag\\
=&\int_{\alpha_m}^{\omega_2}\int_{0}^{\Phi(y)}
 f_{|h_{n}|^2,|h_m|^2}(x,y)dx dy\notag\\
=&\hat{c}_{mn}\!\!\sum_{p=0}^{m-n-\!1}\!\!\hat{c}_p\!\sum_{l=0}^{n-1}\hat{c}_l\frac{1}{l+p+1} \times \notag\\ &\Big(\frac{e^{-(M-n-p)\alpha_m}-e^{-(M-n-p)\omega_2}}{M-n-p}   \notag\\
 & -e^{\frac{l+p+1}{\beta\rho_n}}\frac{e^{-t\alpha_m}-e^{-t\omega_2}}{t} \Big).
\end{align}

Therefore, for the case $\frac{\rho_n}{\rho_m}<\kappa_1$,
$P_{1,1}$ can be expressed as (\ref{sum11}), and for the case $\frac{\rho_n}{\rho_m}>\kappa_1$,
$P_{1,1}$ can be expressed as (\ref{sum12}).

Finally, by following similar methods above, the expression for $P_{1,2}$, $P_{2,1}$ and $P_{2,2}$ can be obtained and the proof is complete.

\section{Proof for Corollary $4$ ($m>n$)}

By noting that  the approximation procedures for $P_1$, $P_{2,1}$, and $P_{2,2}$ are similar,
the next section will only provide the proof for $P_1$ when $\frac{\rho_n}{\rho_m}\le \kappa_1$, due to limited space.

First, recall that $P_1$ is the sum of $P_{1,1}$ and $P_{1,2}$,
in the following,
the approximation for $P_{1,1}$ is obtained first, and then the
approximation for $P_{1,2}$ is developed.

Recall that when $\frac{\rho_n}{\rho_m}\le \kappa_1$, $P_{1,1}$ can be expressed as follows:
\begin{align}
P_{1,1}=&P\left(|h_n|^2\!\le\! \Phi(|h_m|^2),\alpha_m\!<\!|h_m|^2\!<\! \omega_1\right)
\notag\\
&+ P\left(|h_n|^2\!\le\! |h_m|^2 ,\omega_1\!<\!|h_m|^2\!<\! \omega_2\right).
\end{align}

 In order to facilitate approximation, the pdf of $|h_n|^2$ and $|h_m|^2$ ($m>n$) can be rewritten as follows:
 \begin{align}
&f_{|h_{n}|^2,|h_m|^2}(x,y)\\
=&\hat{c}_{mn}\!\left( F(x) \right)^{n\!-\!1}\!\!\left(  F(y) \!-\!F(x) \right)^{m\!-\!n\!-\!1}\!\!\left( 1\!-\!F(y) \right)^{M\!-\!m}\!\!f(x)f(y)\notag\\
= & \hat{c}_{mn}\!\sum\limits_{p=0}^{m\!-\!n\!-\!1}\!\!\!
\left(\!\!\begin{array}{c}
		m\!-\!n\!-\!1\\
		p
	\end{array}\!\!\right)\!(-1)^p f(y) \notag \\
&\times \left(F(y)\right)^{m-n-1-p}\left(1-F(y)\right)^{M-m}f(x)\left(F(x)\right)^{n-1+p},\notag
 \end{align}
where $x<y$, and the last step is obtained by applying the binomial expansion.

Therefore, when $\frac{\rho_n}{\rho_m}\le \kappa_1$, $P_{1,1}$ can be rewritten as follows:
\begin{align}
P_{1,1}
=&\!\!\underbrace{\int_{\alpha_m}^{\omega_1}\!\!\!\!\int_{0}^{\frac{y\alpha_m^{-1}-1}{\beta\rho_n}}f_{|h_{n}|^2,|h_m|^2}(x,y)dx dy}_{\Gamma_3}\notag\\
&
{\color{black}+}\!\!\underbrace{\int_{\omega_1}^{\omega_2}\!\!\!\!\int_{0}^{y}f_{|h_{n}|^2,|h_m|^2}(x,y)dx dy}_{\Gamma_4}.
\end{align} 
By using the joint pdf, the first term of $P_{1,1}$, $\Gamma_3$, can be evaluated as follows:
\begin{align}
 &\Gamma_3=\hat{c}_{mn}\!\sum\limits_{p=0}^{m\!-\!n\!-\!1}\!\!\!
\left(\!\!\begin{array}{c}
		m\!-\!n\!-\!1\\
		p
	\end{array}\!\!\right)\!(-1)^p \int_{\alpha_m}^{\omega_1}\!\!\!\!\int_{0}^{\frac{y\alpha_m^{-1}-1}{\beta\rho_n}}\!\!\!\!f(y)\\
&\times \left(F(y)\right)^{m-n-1-p}\!\left(1-F(y)\right)^{M-m}\!\!f(x)\left(F(x)\right)^{n-1+p}dx dy \notag\\
&=\hat{c}_{mn}\!\sum\limits_{p=0}^{m\!-\!n\!-\!1}\!\!\!
\left(\!\!\begin{array}{c}
		m\!-\!n\!-\!1\\
		p
	\end{array}\!\!\right)\!\frac{(-1)^p }{n+p} \int_{\frac{\epsilon_m}{\rho_m}}^{\frac{\varpi_1}{\rho_m}}\!\!\!\!f(y)\left(F(y)\right)^{m-n-1-p}\!\!\times\notag\\
&\left(1\!-\!F(y)\right)^{\!M\!-\!m}\!\!\Big(\Big(F(\frac{y\alpha_m^{-1}-1}{\beta\rho_n})\Big)^{n+p}\!\!-\!\left(F(0)\right)^{n+p}\Big)dy\notag.
\end{align}
By applying Taylor series, i.e., $f(x)=e^{-x}\approx 1$ and $F(x)=1-e^{-x} \approx x$ when $x\to 0$, $\Gamma_3$ can be approximated as follows:
\begin{align}
 \Gamma_3\approx &\hat{c}_{mn}\!\sum\limits_{p=0}^{m\!-\!n\!-\!1}\!\!\!
\left(\!\!\begin{array}{c}
		m\!-\!n\!-\!1\\
		p
	\end{array}\!\!\right)\!\frac{(-1)^p }{n+p} \int_{\frac{\epsilon_m}{\rho_m}}^{\frac{\varpi_1}{\rho_m}}\!\!\!\!y^{m-n-1-p}\times\notag\\
&\Big(\Big(\frac{y\alpha_m^{-1}-1}{\beta\rho_n}\Big)^{n+p}\!\!-0\Big) dy  \notag    \\
= & \hat{c}_{mn}\!\sum\limits_{p=0}^{m\!-\!n\!-\!1}\!\!\!
\left(\!\!\begin{array}{c}
		m\!-\!n\!-\!1\\
		p
	\end{array}\!\!\right)\!\frac{(-1)^p }{n+p} \frac{1}{(\beta\eta\rho_m)^{n+p}} \times
\notag \\&
\sum\limits_{q=0}^{n+p} \!\!
\left(\!\!\begin{array}{c}
		n+p\\
		q
	\end{array}\!\!\right)(-1)^q\alpha_m^{-(n+p-q)}\!\!\! \int_{\frac{\epsilon_m}{\rho_m}}^{\frac{\varpi_1}{\rho_m}}\!\!\!y^{m-q-1}dy.
\end{align}
After some algebraic manipulations, $\Gamma_3$ can be further approximated as follows:
 \begin{align}\label{Gamma_3}
 \Gamma_3\approx & \frac{1}{\rho_m^m}\hat{c}_{mn}\!\sum\limits_{p=0}^{m\!-\!n\!-\!1}\!\!\!
\left(\!\!\begin{array}{c}
		m\!-\!n\!-\!1\\
		p
	\end{array}\!\!\right)\!\frac{(-1)^p }{n+p}\notag \\
&\times \sum\limits_{q=0}^{n+p} \!\!
\left(\!\!\begin{array}{c}
		n+p\\
		q
	\end{array}\!\!\right)(-1)^q \frac{\varpi_1^{m-q}-\epsilon_m^{m-q}}{(\beta\eta)^{n+p}\epsilon_m^{n+p-q}(m-q)} .
\end{align}
Similar to the derivation steps for $\Gamma_3$, the high SNR approximation for $\Gamma_4$ can be given by
 \begin{align}\label{Gamma_4}
 \Gamma_4\approx & \frac{1}{\rho_m^m}\hat{c}_{mn}\!\sum\limits_{p=0}^{m\!-\!n\!-\!1}\!\!\!
\left(\!\!\begin{array}{c}
		m\!-\!n\!-\!1\\
		p
	\end{array}\!\!\right)\!(-1)^p\frac{\varpi_2^m-\varpi_1^m }{(n+p)m}.
\end{align}
Therefore, when $\frac{\rho_n}{\rho_m}\le \kappa_1$, $P_{1,1}$ can be approximated as follows:
\begin{align}\label{E_P1,1}
P_{1,1} \approx & \frac{1}{\rho_m^m}\hat{c}_{mn}\!\!\!\sum\limits_{p=0}^{m\!-\!n\!-\!1}\!\!\!
\left(\!\!\begin{array}{c}
		\!m\!-\!n\!-\!1\!\\
		p
	\end{array}\!\!\right)\!\frac{(-1)^p }{n+p} \left({\color{black}\frac{\varpi_2^m\!-\!\varpi_1^m }{m}}\right.\\
&\left.+\sum\limits_{q=0}^{n+p} \!\!
\left(\!\!\begin{array}{c}
		n+p\\
		q
	\end{array}\!\!\right)(-1)^q \frac{\varpi_1^{m-q}-\epsilon_m^{m-q}}{(\beta\eta)^{n+p}\epsilon_m^{n+p-q}(m-q)}\right).\notag
\end{align}

For $P_{1,2}$, when $\frac{\rho_n}{\rho_m} \le \kappa_1$, it can be rewritten as follows:
\begin{align}
P_{1,2}=&\underbrace {P\left(|h_n|^2\!\le\! |h_m|^2,\omega_2\!<\!|h_m|^2\!<\! \omega_3\right)}_{\Gamma_5}
\notag\\
&+ \underbrace{P\left(|h_n|^2\!\le\! \omega_3 ,|h_m|^2\!>\! \omega_3\right)}_{\Gamma_6}
\end{align}
Similar to the derivation steps above, $\Gamma_5$ can be approximated as follows:
\begin{align}
\Gamma_5 \approx & \label{Gamma_5} \frac{1}{\rho_m^m}\hat{c}_{mn}\!\sum\limits_{p=0}^{m\!-\!n\!-\!1}\!\!\!
\left(\!\!\begin{array}{c}
		m\!-\!n\!-\!1\\
		p
	\end{array}\!\!\right)\!(-1)^p\frac{\varpi_3^m-\varpi_2^m }{(n+p)m}.
\end{align}
To facilitate the calculation of the approximate results,
$\Gamma_6$ can be rewritten as follows:
\begin{align}
\Gamma_6 = & P\left( |h_n|^2\le \omega_3  \right)-P\left( |h_n|^2\le \omega_3,|h_m|^2\le \omega_3  \right)
\end{align}
$P\left( |h_n|^2\le \omega_3  \right)$ is a function of one order statistics, $|h_n|^2$, whose marginal pdf is given by
\begin{align}
f_{|h_n|^2}(x)=\hat{c}_n
{\color{black}\left( F(x) \right)^{n-1}}f(x)^{M-n+1}.
\end{align}
By applying the above density functions, $\Gamma_6$ can be expressed as follows:
\begin{align}
 \Gamma_6= \int_{0}^{\omega_3}\!\!\!f_{|h_{n}|^2}(x)dx-
\!\!\int_{0}^{\omega_3}\!\!\!\!\int_{0}^{\omega_3}\!\!f_{|h_{n}|^2,|h_m|^2}(x,y)dx dy
\end{align}
Similar to the derivation steps above, $\Gamma_6$ can be approximated as follows:
\begin{align}\label{Gamma_6}
\Gamma_6 \approx & \frac{1}{\rho_m^n}\hat{c}_n\frac{\varpi_3^n}{n}\\
&-\frac{1}{\rho_m^m}\hat{c}_{mn}\!\sum\limits_{p=0}^{m\!-\!n\!-\!1}\!\!\!
\left(\!\!\begin{array}{c}
		m\!-\!n\!-\!1\\
		p
	\end{array}\!\!\right)\!\frac{(-1)^p\varpi_3^m }{(m-n-p)(n+p)}.\notag
\end{align}
Therefore, when $\frac{\rho_n}{\rho_m}\le \kappa_1$, $P_{1,2}$ can be approximated as follows:
\begin{align}\label{E_P1,2}
P_{1,2} \approx & \frac{1}{\rho_m^n}\hat{c}_n\frac{\varpi_3^n}{n}
{\color{black}+}\frac{1}{\rho_m^m}\hat{c}_{mn}\!\sum\limits_{p=0}^{m\!-\!n\!-\!1}\!\!\!
\left(\!\!\begin{array}{c}
		m\!-\!n\!-\!1\\
		p
	\end{array}\!\!\right)\!\frac{(-1)^p}{n+p}\notag\\
&\times\left(\frac{\varpi_3^m-\varpi_2^m }{m}-\frac{\varpi_3^m }{m-n-p}\right).
\end{align}

Therefore, when $\frac{\rho_n}{\rho_m}\le \kappa_1$, $P_1$ can be approximated as the sum of (\ref{E_P1,1}) and (\ref{E_P1,2}).

By following the same method for the approximation above, the approximations for $P_1$ when $\frac{\rho_n}{\rho_m}> \kappa_1$, $P_{2,1}$ and $P_{2,2}$ can be obtained. Hence, the proof is complete.

\bibliographystyle{IEEEtran}
\bibliography{IEEEabrv,ref}

\begin{thebibliography}{10}
\providecommand{\url}[1]{#1}
\csname url@samestyle\endcsname
\providecommand{\newblock}{\relax}
\providecommand{\bibinfo}[2]{#2}
\providecommand{\BIBentrySTDinterwordspacing}{\spaceskip=0pt\relax}
\providecommand{\BIBentryALTinterwordstretchfactor}{4}
\providecommand{\BIBentryALTinterwordspacing}{\spaceskip=\fontdimen2\font plus
\BIBentryALTinterwordstretchfactor\fontdimen3\font minus \fontdimen4\font\relax}
\providecommand{\BIBforeignlanguage}[2]{{%
\expandafter\ifx\csname l@#1\endcsname\relax
\typeout{** WARNING: IEEEtran.bst: No hyphenation pattern has been}%
\typeout{** loaded for the language `#1'. Using the pattern for}%
\typeout{** the default language instead.}%
\else
\language=\csname l@#1\endcsname
\fi
#2}}
\providecommand{\BIBdecl}{\relax}
\BIBdecl

\bibitem{you2021towards}
X.~You, C.-X. Wang, J.~Huang, X.~Gao, Z.~Zhang, M.~Wang, Y.~Huang, C.~Zhang, Y.~Jiang, J.~Wang \emph{et~al.}, ``{Towards 6G wireless communication networks: Vision, enabling technologies, and new paradigm shifts},'' \emph{{Sci. China Inf. Sci. }}, vol.~64, pp. 1--74, Feb. 2021.

\bibitem{3GPPNOMAR16}
{3GPP, TR38.812 V16.0.0}, ``{Study on Non-Orthogonal Multiple Access (NOMA) for NR},'' Dec. 2018.

\bibitem{liu2022evolution}
Y.~Liu, S.~Zhang, X.~Mu, D.~Zhiguo, R.~Schober, N.~Al-Dhahir, and E.~Hossain, ``{Evolution of NOMA toward next generation multiple access (NGMA) for 6G},'' \emph{{IEEE} J. Sel. Areas Commun.}, vol.~40, no.~4, pp. 1037--1071, Jan. 2022.

\bibitem{recommendation2023framework}
{I. T. U. (ITU)}, ``{Framework and overall objectives of the future development of IMT for 2030 and beyond},'' Nov. 2023, {recommendation ITU-R M.2160-0}.

\bibitem{sun2024study}
L.~Sun, Z.~Zhao, S.~Wang, Z.~Ding, and M.~Peng, ``{On the Study of Non-Orthogonal Multiple Access (NOMA)-Assisted Integrated Sensing and Communication (ISAC)},'' \emph{IEEE Trans. Commun.}, 2024, early access, DOI: 10.1109/TCOMM.2024.3407202.

\bibitem{new2023fluid}
W.~K. New, K.-K. Wong, H.~Xu, K.-F. Tong, C.-B. Chae, and Y.~Zhang, ``{Fluid antenna system enhancing orthogonal and non-orthogonal multiple access},'' \emph{IEEE Commun. Lett.}, vol.~28, no.~1, pp. 218--222, Jan. 2024.

\bibitem{li2023achievable}
Q.~Li, M.~El-Hajjar, Y.~Sun, I.~Hemadeh, A.~Shojaeifard, Y.~Liu, and L.~Hanzo, ``{Achievable rate analysis of the STAR-RIS-aided NOMA uplink in the face of imperfect CSI and hardware impairments},'' \emph{IEEE Trans. Commun.}, vol.~71, no.~10, pp. 6100--6114, Oct. 2023.

\bibitem{zhu2020power}
J.~Zhu, Y.~Huang, J.~Wang, K.~Navaie, and Z.~Ding, ``Power efficient irs-assisted noma,'' \emph{IEEE Transactions on Communications}, vol.~69, no.~2, pp. 900--913, Feb. 2020.

\bibitem{mu2023exploiting}
X.~Mu and Y.~Liu, ``{Exploiting semantic communication for non-orthogonal multiple access},'' \emph{{IEEE} J. Select. Areas Commun.}, vol.~41, no.~8, pp. 2563--2576, Aug. 2023.

\bibitem{chu2020robust}
J.~Chu, X.~Chen, C.~Zhong, and Z.~Zhang, ``{Robust design for NOMA-based multibeam LEO satellite Internet of Things},'' \emph{IEEE Internet of Things J.}, vol.~8, no.~3, pp. 1959--1970, Feb. 2020.

\bibitem{Ding2019MEC}
Z.~{Ding}, P.~{Fan}, and H.~V. {Poor}, ``{Impact of non-orthogonal multiple access on the offloading of mobile edge computing},'' \emph{{IEEE} Trans. Commun.}, vol.~67, no.~1, pp. 375--390, Jan. 2019.

\bibitem{ding2024backcom}
Z.~Ding, ``{BackCom Assisted Hybrid NOMA Uplink Transmission for Ambient IoT},'' 2024, {arXiv preprint arXiv:2403.16498}.

\bibitem{ding2022hybrid}
Z.~Ding, D.~Xu, R.~Schober, and H.~V. Poor, ``Hybrid noma offloading in multi-user mec networks,'' \emph{IEEE Trans. Wireless Commun.}, vol.~21, no.~7, pp. 5377--5391, Jul. 2022.

\bibitem{liu2021latency}
L.~Liu, B.~Sun, Y.~Wu, and D.~H. Tsang, ``{Latency optimization for computation offloading with hybrid NOMA--OMA transmission},'' \emph{IEEE Internet of Things J.}, vol.~8, no.~8, pp. 6677--6691, Aug. 2021.

\bibitem{wei2022energy}
X.~Wei, H.~Al-Obiedollah, K.~Cumanan, Z.~Ding, and O.~A. Dobre, ``{Energy efficiency maximization for hybrid TDMA-NOMA system with opportunistic time assignment},'' \emph{{{IEEE} Trans. Veh. Technol.}}, vol.~71, no.~8, pp. 8561--8573, Aug. 2022.

\bibitem{ding2024utilizing}
Z.~Ding and H.~V. Poor, ``{Utilizing imperfect resolution of near-field beamforming: A hybrid-NOMA perspective},'' \emph{IEEE Commun. Lett.}, 2024, {early access DOI: 10.1109/LCOMM.2024.3398412}.

\bibitem{ding2024design}
Z.~Ding, R.~Schober, and H.~V. Poor, ``{Design of downlink hybrid NOMA transmission},'' 2024, {arXiv preprint arXiv:2401.16965}.

\bibitem{yu2022irs}
J.~Yu, Y.~Li, X.~Liu, B.~Sun, Y.~Wu, and D.~H.-K. Tsang, ``{IRS assisted NOMA aided mobile edge computing with queue stability: Heterogeneous multi-agent reinforcement learning},'' \emph{IEEE Trans. Wireless Commun.}, vol.~22, no.~7, pp. 4296--4312, Jul. 2022.

\bibitem{wang2022reinforcement}
K.~Wang, H.~Li, Z.~Ding, and P.~Xiao, ``{Reinforcement learning based latency minimization in secure NOMA-MEC systems with hybrid SIC},'' \emph{IEEE Trans. Wireless Commun.}, vol.~22, no.~1, pp. 408--422, Jan. 2022.

\bibitem{chaieb2022deep}
C.~Chaieb, F.~Abdelkefi, and W.~Ajib, ``{Deep reinforcement learning for resource allocation in multi-band and hybrid OMA-NOMA wireless networks},'' \emph{IEEE Trans. Commun.}, vol.~71, no.~1, pp. 187--198, Jan. 2022.

\bibitem{higuchi2013non}
K.~Higuchi and Y.~Kishiyama, ``{Non-orthogonal access with random beamforming and intra-beam SIC for cellular MIMO downlink},'' in \emph{IEEE Veh. Tech. Conf.,}, Las Vegas, NV, US, Sep. 2013, pp. 1--5.

\bibitem{gao2017theoretical}
Y.~{Gao}, B.~{Xia}, K.~{Xiao}, Z.~{Chen}, X.~{Li}, and S.~{Zhang}, ``Theoretical analysis of the dynamic decode ordering sic receiver for uplink noma systems,'' \emph{{IEEE} Commun. Lett.}, vol.~21, no.~10, pp. 2246--2249, Jun. 2017.

\bibitem{Xia2018outage}
B.~{Xia}, J.~{Wang}, K.~{Xiao}, Y.~{Gao}, Y.~{Yao}, and S.~{Ma}, ``{Outage Performance Analysis for the Advanced SIC Receiver in Wireless NOMA Systems},'' \emph{{IEEE} Trans. Veh. Technol.}, vol.~67, no.~7, pp. 6711--6715, Mar. 2018.

\bibitem{zhou2018state}
F.~Zhou, Y.~Wu, Y.-C. Liang, Z.~Li, Y.~Wang, and K.-K. Wong, ``{State of the art, taxonomy, and open issues on cognitive radio networks with NOMA},'' \emph{IEEE Wireless Commun.}, vol.~25, no.~2, pp. 100--108, 2018.

\bibitem{Dhakal2019noma}
S.~{Dhakal}, P.~A. {Martin}, and P.~J. {Smith}, ``{NOMA With Guaranteed Weak User QoS: Design and Analysis},'' \emph{IEEE Access}, vol.~7, pp. 32\,884--32\,896, Mar. 2019.

\bibitem{Ding2019simple}
Z.~{Ding}, R.~{Schober}, P.~{Fan}, and H.~V. {Poor}, ``{Simple Semi-Grant-Free Transmission Strategies Assisted by Non-Orthogonal Multiple Access},'' \emph{IEEE Trans. Commun.}, vol.~67, no.~6, pp. 4464--4478, Mar. 2019.

\bibitem{ding2021new}
Z.~Ding, R.~Schober, and H.~V. Poor, ``{A new QoS-guarantee strategy for NOMA assisted semi-grant-free transmission},'' \emph{IEEE Trans. Commun.}, vol.~69, no.~11, pp. 7489--7503, Jul. 2021.

\bibitem{sun2021new}
Y.~Sun, Z.~Ding, and X.~Dai, ``{A new design of hybrid SIC for improving transmission robustness in uplink NOMA},'' \emph{{IEEE} Trans. Veh. Technol.}, vol.~70, no.~5, pp. 5083--5087, Mar. 2021.

\bibitem{lu2022advanced}
H.~Lu, X.~Xie, Z.~Shi, H.~Lei, H.~Yang, and J.~Cai, ``{Advanced NOMA assisted semi-grant-free transmission schemes for randomly distributed users},'' \emph{IEEE Trans. Wireless Commun.}, vol.~22, no.~7, pp. 4638--4653, Jul. 2022.

\bibitem{sun2023hybrid}
Y.~Sun, W.~Cao, M.~Zhou, and Z.~Ding, ``{Hybrid successive interference cancellation and power adaptation: a win-win wtrategy for robust uplink nOMA transmission},'' \emph{IEEE Trans. Commun.}, vol.~72, no.~2, pp. 771--785, Feb. 2024.

\end{thebibliography}
\end{document}